\documentclass[a4paper,aps,prd,twocolumn,eqsecnum,amsmath,amssymb,nofootinbib,superscriptaddress]{revtex4-2}
\usepackage{dcolumn}
\usepackage{bm}
\usepackage{graphics}
\usepackage{graphicx}
\usepackage{epsfig}
\usepackage{booktabs,multirow}
\usepackage{hhline}
\usepackage{slashed}
\usepackage{rccol}
\usepackage{mathrsfs}
\usepackage{amsmath, amssymb}
\usepackage[dvipsnames]{xcolor,colortbl}
\usepackage{subfigure}
\graphicspath{{figures/}}

\usepackage{xcolor}\colorlet{linkequation}{blue}
\usepackage[colorlinks=true, % false: boxed links; true: colored links
linkcolor=Red,   % color of internal links
filecolor=Red,   % color of file links
citecolor=Green, % color of links to bibliography
urlcolor=Blue,
hyperfootnotes=true]{hyperref}

\newcommand*{\SavedEqref}{}
\let\SavedEqref\eqref
\renewcommand*{\eqref}[1]{%
	\begingroup
	\hypersetup{
		linkcolor=linkequation,
		linkbordercolor=linkequation,
	}%
	\SavedEqref{#1}%
	\endgroup
}

\begin{document}
\title{Solar Neutrino Probes of Light New Physics: Updated Limits from LUX-ZEPLIN Experiment}
%%Begin Author List
\newcommand{\ktu}{Department of Physics,
Karadeniz Technical University, Trabzon, TR61080, Türkiye}
\newcommand{\ktufb}{Graduate School of Natural and Applied Science, Karadeniz Technical University, Trabzon, TR61080, Türkiye}
\author{Mehmet~Demirci}
\email{mehmetdemirci@ktu.edu.tr}
\affiliation{\ktu}

\author{M.~Fauzi~Mustamin}
%\email{mfmustamin@ktu.edu.tr}
\affiliation{\ktu}
\affiliation{\ktufb}
%%End Author List

\date{\today}

\begin{abstract}
The recent low-energy electron recoil (ER) results reported by the LUX-ZEPLIN (LZ) experiment have established the most stringent constraints to date on new physics scenarios, specifically for solar axion-like particles with keV-scale masses and mirror dark matter. 
Motivated by this enhanced sensitivity and the resulting restrictive limits, our present work focuses on probing light mediator models via elastic neutrino-electron scattering induced by solar neutrinos. 
We specifically consider two broad classes of new physics scenarios: (i) universal light mediator models consistent with Lorentz invariance, including scalar, vector, and tensor interactions, and (ii) anomaly-free leptophilic $U(1)'$ gauge extensions featuring a new vector mediator associated with the $L_e-L_\mu$, $L_e-L_\tau$, $L_\mu-L_\tau$, and $L_e+2L_\mu+2L_\tau$ symmetries. 
By incorporating contributions from these interactions into the neutrino-electron scattering cross-section and utilizing the most precise solar neutrino flux predictions, we analyze the latest LZ ER datasets. We report novel constraints on the coupling-mass parameter space for these models. 
Furthermore, we contextualize our findings by comparing them with established bounds from various laboratory, cosmological, and astrophysical sources.
Our analysis demonstrates that LZ data provide significantly improved limits in previously unconstrained regions of the parameter space.
\end{abstract}
%\medskip

\maketitle
\section{Introduction}
Solar neutrinos represent one of the most abundant sources of natural neutrinos reaching Earth. Originating as electron neutrinos from fusion reactions in the solar core, they undergo flavor transitions during propagation \cite{Davis:1968cp}. The detection of solar neutrinos has played a central role in establishing neutrino oscillations and the matter effect through a variety of experimental channels, including charged current (CC) interactions \cite{GALLEX:1992gcp, Cleveland:1998nv, SAGE:1999uje}, neutral current (NC) interactions \cite{SNO:2002tuh, SNO:2003bmh, SNO:2008gqy}, and elastic scattering \cite{Kamiokande-II:1989hkh, Super-Kamiokande:2001ljr, Borexino:2007kvk, Borexino:2008fkj}. Among these, elastic neutrino--electron scattering (E$\nu$ES) is a purely leptonic process within the Standard Model (SM), mediated by both CC and NC interactions, in which neutrinos scatter off atomic electrons \cite{tHooft:1971ucy}. Owing to its clean experimental signature, this process provides a particularly robust probe of the SM electroweak sector. For solar neutrinos, the recoil electrons are predominantly emitted in the forward direction along the Sun--Earth axis \cite{Bahcall:1995mm}, thereby preserving sensitivity to multiple neutrino flavors.

The experimental observation of solar neutrino interactions has become achievable within modern Dark Matter (DM) Direct Detection (DD) experiments.
Although originally designed to detect DM candidates through nuclear recoils (NRs) \cite{Drukier:1984vhf, Goodman:1984dc, Drukier:1986tm}, these detectors have now achieved remarkable sensitivity to electron recoil (ER) events as well. Consequently, the elastic scattering of solar neutrinos off electrons constitutes not only an irreducible background---often referred to as the ``neutrino floor''---but also a powerful signal channel for probing new physics, yielding observable event rates at current experiments \cite{Cerdeno:2016sfi, Schwemberger:2022fjl}. 
This capability is primarily enabled by the dual-phase liquid xenon (LXe) time projection chamber (TPC) technology, which combines an ultra-low energy threshold, scalability to multi-ton target masses, and extremely low background levels. These features provide broad sensitivity to a wide variety of new physics scenarios beyond conventional searches for dark matter in the form of weakly interacting massive particles (WIMPs).
%WIMP dark matter searches.

At present, the large fiducial volumes of LXe detectors are operational at XENONnT (5.9~ton) \cite{XENON:2020kmp, XENON:2022ltv}, PandaX-4T (3.7~ton) \cite{PandaX:2014mem, PandaX-II:2017hlx, PandaX:2024cic}, and LUX-ZEPLIN (LZ) (5.5~ton) \cite{LUX:2015abn, LZ:2018qzl, LZ:2022lsv, LZ:2024zvo, LZ:2025zpw}. These experiments have already accumulated exposures of 1.16, 0.63, and 0.90~ton--year, respectively, within relatively short data-taking periods during their initial science runs. The success of the current generation of detectors has further motivated ambitious next-generation projects; while large-scale argon detectors like DarkSide-20k \cite{DARKSIDE20K:2021} prepare to explore new frontiers, the xenon community—represented by the XLZD Consortium (XENON, LZ, and DARWIN)—aims to develop a 60-ton LXe detector in the next decade to push sensitivity limits even further \cite{DARWIN:2020bnc, Aalbers:2022dzr}.

The LZ Experiment operates the largest dual-phase xenon TPC in existence, situated 4850 feet beneath the surface within the Davis Cavern at the Sanford Underground Research Facility (Lead, SD). Benefiting from a rock overburden of 4300 meters water equivalent, the detector focuses on the direct observation of WIMPs via scattering interactions with xenon nuclei.
The LZ Collaboration has recently reported the most sensitive WIMP search to date by combining the 60~live-day exposure from its first science run (WS2022) \cite{LZ:2022lsv} with an additional 220~live-day dataset (WS2024) \cite{LZ:2024zvo}. Two regions of interest (ROIs), a 1D and a 2D ROI, were employed to accommodate different signal topologies. In addition to WIMP searches, the collaboration has derived constraints on several new physics scenarios affecting ER signals \cite{LZ:2025zpw}, including mirror dark matter, absorption of bosonic dark matter candidates, electromagnetic properties of solar neutrinos, and solar axion-like particles (ALPs). These results currently provide the most stringent limits on solar ALPs with keV-scale masses and on mirror dark matter. This remarkable sensitivity clearly demonstrates the strong potential of such experiments to probe new physics through the solar neutrino--induced neutrino--electron scattering process.

Motivated by these developments, in this work we investigate new physics contributions from light mediator models in light of recent ER signals of the LZ experiment. These models generically induce spectral distortions by enhancing the E$\nu$ES differential cross section, especially in the regime of small mediator masses and low recoil energies. 
In many extensions of the SM, low-mass particles emerging from hidden sectors naturally arise and provide well-motivated realizations of such scenarios. 
We focus on two broad classes of models: universal light mediator models \cite{Cirelli:2013ufw, Abdallah:2015ter} and lepton flavor--dependent $U(1)'$ gauge models \cite{Mohapatra:1980qe, He:1991qd}. The former class includes universal scalar, tensor and vector interactions allowed by Lorentz invariance. The latter class consists of anomaly-free leptophilic gauge symmetries $U(1)_{L_e-L_\mu}$, $U(1)_{L_e-L_\tau}$, $U(1)_{L_\mu-L_\tau}$, and $U(1)_{L_e+2L_\mu+2L_\tau}$, mediated by a new vector boson ($Z'$), which couples exclusively to leptons with flavor-dependent charges. 
The corresponding couplings therefore differ from those of the universal vector mediator according to the assigned lepton charges. All of these scenarios are theoretically well motivated and offer complementary explanations for several emerging anomalies and tensions observed in low-energy precision measurements. Using the most recent LZ datasets, we derive robust and, for the first time, direct limits on the coupling--mass parameter space of these light mediator models.

The paper is structured as follows: Section~\ref{sec:nue} details the theoretical formalism for E$\nu$ES, covering both the SM and extensions involving new light mediators. Subsequently, Section~\ref{sec:stat} is dedicated to the statistical framework used to derive the exclusion limits. Section~\ref{sec:resdis} details the expected recoil spectra, the resulting constraints on the model parameters, and a systematic comparison with previously established limits. Finally, Section~\ref{sec:summ} provides an overall summary of our analysis together with the concluding remarks.

\section{Theoretical Framework}\label{sec:nue}
This section establishes the theoretical framework necessary for calculating the expected event rate of solar neutrinos scattering elastically off electrons in the LZ experiment. We present the relevant interaction cross sections within the SM and in the presence of light mediator models in Subsections~\ref{sec:SM} and \ref{sec:BSM}, respectively.

\subsection{E$\nu$ES in the Standard Model}\label{sec:SM}
Within the SM framework, E$\nu$ES represents a fundamental leptonic interaction. It proceeds through two channels: the exchange of a charged W boson (restricted to $\nu_e$) or a neutral Z boson (valid for all flavors). The corresponding differential cross-section as a function of the electron recoil energy $T_{e}$ can be expressed as
\begin{widetext}
	\begin{align} 
		\begin{split}
			\left[\frac{d\sigma_{\nu_\ell}}{dT_{e}}\right]_{\mathrm{SM}} = \frac{G_F^2 m_e}{2\pi} &\Bigg[ (g_V+g_A)^2 + (g_V-g_A)^2 \left(1-\frac{T_{e}}{E_\nu}\right)^2 - (g_V^{2}-g_A^{2}) \frac{m_e T_{e}}{E_\nu^2} \Bigg],
		\end{split}
		\label{eq:sm_nue}
	\end{align}
\end{widetext}
where $m_e$ denotes the electron mass, $E_\nu$ is the incident neutrino energy, and $G_F$ is the Fermi coupling constant.
%$G_F = 1.1663787\times 10^{-5}\,\mathrm{GeV}^{-2}$.  
The vector and axial-vector couplings of neutrinos to electrons are given by
\begin{align}
g_V &= -\frac12 + 2\sin^2\theta_W + \delta_{\ell e},\\
g_A &= -\frac12 + \delta_{\ell e},
\end{align}
where $\delta_{\ell e}$ is the Kronecker delta accounting for the charged--current contribution that appears only for electron neutrinos.  
The low-energy $\overline{\text{MS}}$ value of the weak mixing angle is given by $\sin^2\theta_W = 0.23873$~\cite{ParticleDataGroup:2024cfk}.
The couplings vary according to the neutrino's flavor species, denoted by $\ell = e, \mu, \tau$.  
Including radiative electroweak corrections, we take the effective couplings as ~\cite{Erler:2013c, AtzoriCorona:2025ygn}
\begin{align}
g_V^{\nu_e e} &= 0.9524, &\qquad g_A^{\nu_e e} &= 0.4938,\\
g_V^{\nu_\mu e} &= -0.0394, &\qquad g_A^{\nu_\mu e} &= -0.5062,\\
g_V^{\nu_\tau e} &= -0.0350, &\qquad g_A^{\nu_\tau e} &= -0.5062.
\end{align}

The neutrino-electron scattering process is characterized by high directionality, where the outgoing electron is kinematically constrained to align closely with the incoming neutrino's path~\cite{Formaggio:2012cpf}. This signature was successfully exploited by major neutrino observatories like SNO~\cite{SNO:2002tuh, SNO:2003bmh, SNO:2008gqy}, Super-Kamiokande~\cite{Kamiokande-II:1989hkh, Super-Kamiokande:2001ljr}, and BOREXINO~\cite{Borexino:2007kvk, Borexino:2008fkj} to identify solar neutrinos. However, in DD experiments, these interactions with atomic electrons act as a pervasive background alongside CE$\nu$NS. Although discrimination is possible in principle, practical limitations in background removal result in a significant degeneracy with dark matter-induced electron recoils.

\subsection{New Light Mediator Models}\label{sec:BSM}
At low-energy neutrino experiments, elastic neutrino-electron scattering serves as a sensitive probe for new physics, particularly in scenarios involving light mediators coupled to SM leptons. In this work, we investigate universal light mediator models~\cite{Cirelli:2013ufw, Abdallah:2015ter}, which encompass scalar, vector, and tensor interactions consistent with Lorentz invariance. Additionally, we examine anomaly-free $U(1)'$ extensions of the SM featuring new vector mediators, specifically the $L_e-L_\mu$, $L_e-L_\tau$, $L_\mu-L_\tau$, and $L_e+2L_\mu+2L_\tau$ gauge models~\cite{Mohapatra:1980qe, He:1991qd}, where charges are exclusively leptonic. These simplified frameworks are constructed with a minimal number of new particles and interactions, serving as effective low-energy limits of more comprehensive beyond the Standard Model (BSM) theories. By reducing the parameter space to a few key variables---namely coupling constants and masses---such models allow us to explore an extensive spectrum of new physics signatures without the need to specify a full high-energy UV-complete theory.

\begin{figure}[!h]
	\centering
	\includegraphics[scale=0.65]{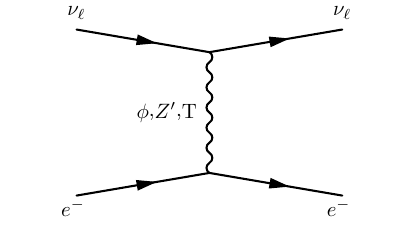}
	\caption{Feynman diagram for E$\nu$ES mediated by the new mediators $\phi$, $Z'$, or $\mathrm{T}$.}
	\label{fig:nue_diag_bsm}
\end{figure}
In the presence of the light mediators considered in this study, the E$\nu$ES process is illustrated schematically in Fig.~\ref{fig:nue_diag_bsm}. We evaluate the contributions of scalar, vector, tensor, and lepton-flavor-dependent $U(1)'$ gauge bosons with mass $m_X$, characterized by an effective coupling 
\begin{align}
g_X = \sqrt{g_X^{e}\, g_X^{\nu_\ell}},
\end{align}
where $g_X^{e}$ and $g_X^{\nu_\ell}$ are the corresponding couplings to electrons and neutrinos.  
Our aim is to determine how sensitively these interactions can be probed using the most recent electron recoil data from dark matter DD experiments.
The low-threshold measurements of the LZ Experiment, in particular, provide an excellent environment to test such scenarios. For mediator masses comparable to the momentum transfer relevant to solar neutrino scattering, the new interaction can noticeably distort the recoil energy spectrum, thereby yielding competitive and, in certain regions, substantially stronger constraints in the $(m_X, g_X)$ parameter space.

\subsubsection{Universal scalar}
We begin by examining a light scalar mediator $\phi$ that couples in a universal manner to all charged leptons and neutrino flavors in the SM.  Such an interaction introduces an additional scalar current in E$\nu$ES processes.  
The corresponding interaction Lagrangian is expressed as \cite{Cerdeno:2016sfi}

\begin{align}
\begin{split}
\mathcal{L}_\phi \supset -  \phi &\biggl[g_{\phi}^e \bar{e} e + g_{\phi}^{\nu_\ell} \bar{\nu_\ell}_R {\nu_\ell}_L  + h.c. \biggr],
%-\frac{1}{2} m_\phi^2 \phi^2,
\end{split}
\end{align}
where the scalar coupling constant $g_{\phi}^e$ is for electron and $g_{\phi}^{\nu_\ell}$ for neutrino. 
In this framework, the additional contribution of the scalar mediator to the E$\nu$ES process is given by \cite{Cerdeno:2016sfi}
\begin{align}
\biggl[\frac{d\sigma_{\nu_\ell}}{dT_{e}}\biggr]_\phi &=
\frac{(g_\phi^e g_\phi^{\nu_\ell})^2}{4\pi\left(m_\phi^2 + 2m_e T_{e}\right)^2}
\left[\frac{T_e\, m_e^2}{E_\nu^{\,2}}\right],
\label{eq:difcross_scalar}
\end{align}
where the expression holds for an incident neutrino of flavor $\ell$. 
Since the scalar interaction generates a distinct Lorentz structure relative to the SM amplitudes, it does not interfere with the SM contributions.  Consequently, the total E$\nu$ES differential cross-section is obtained by a simple incoherent addition of the scalar term on top of the SM prediction.

\subsubsection{Universal vector}
We next examine a light vector mediator, denoted by $Z'$, which couples in a universal manner to SM electrons and neutrinos. The associated interaction Lagrangian takes the form
\begin{align}
\begin{split}
\mathcal{L}_{Z'} \supset Z'_\mu & \biggl[Q^e_{Z'} g_{Z'}^e \bar{e} \gamma^\mu e + Q^{\nu_\ell}_{Z'} g_{Z'}^{\nu_\ell} \bar{\nu_\ell}_L \gamma^\mu  {\nu_\ell}_L  \biggr],
%+\frac{1}{2} m_{Z'} Z'^{\mu} Z'_\mu
\end{split}
\end{align}
where the vector coupling constants $g_{Z'}^e$ and $g_{Z'}^{\nu_\ell}$ are for electron and neutrino, respectively. The vector gauge charges of electrons and neutrinos are represented by $Q^{e}_{Z'}$ and $Q^{\nu_\ell}_{Z'}$, respectively.  
These charge assignments allow us to treat the interaction in a model-independent way and can be embedded into a broad class of anomaly-free, UV-complete $U(1)'$ extensions of the SM that contain only the SM fields supplemented by right-handed neutrinos \cite{Allanach:2019}.

The contribution of a $Z'$ mediator to the E$\nu$ES cross section can be incorporated by performing the replacement \cite{Ballet:2019}
\begin{align}
    g_V \;\longrightarrow\;
    g_V +
    \frac{Q^{e}_{Z'}\,g_{Z'}^{e}\,Q^{\nu_\ell}_{Z'}\,g_{Z'}^{\nu_\ell}}
         {2\sqrt{2}\,G_F\,(m_{Z'}^{2}+2m_e T_e)}
    \label{eq:vec_chrg}
\end{align}
in Eq.~\eqref{eq:sm_nue}.  
This expression remains valid across both the light- and heavy-mediator regimes, noting that typical momentum transfers in direct detection experiments are $\mathcal{O}(100~\text{keV})$.

For the universal vector scenario, one has $Q^{e}_{Z'} = Q^{\nu_\ell}_{Z'} = 1$.  
With this choice, the resulting cross section closely resembles that of the well-studied $U(1)_{B-L}$ model, where the corresponding charges satisfy $Q^{e}_{Z'} = Q^{\nu_\ell}_{Z'} = -1$.  
The distinction between these two models becomes apparent in CE$\nu$NS processes, in which quark charges contribute and differ by an overall factor of $1/3$ \cite{DeRomeri:2022twg, Demirci:2023tui}.  
However, for E$\nu$ES, where only leptonic currents are involved, both models yield identical predictions.

\subsubsection{Universal tensor}

We next examine a light tensor mediator $\mathrm{T}$ that couples universally to SM electrons and neutrinos.  
The corresponding interaction Lagrangian can be written as \cite{Barranco:2011wx}
\begin{align}
\begin{split}
\mathcal{L}_{\mathrm{T}} \supset &\biggl[ g_\mathrm{T}^{e} \bar{e} \sigma^{\mu\nu} e -  g_\mathrm{T}^{\nu_l} \bar{\nu_\ell}_R \sigma^{\mu\nu} {\nu_\ell}_L\biggr] \mathrm{T}_{\mu\nu},
\end{split}
\end{align}
where $\sigma_{\mu\nu} = i(\gamma_\mu\gamma_\nu - \gamma_\nu\gamma_\mu)/2$.  
Here, $g_{\mathrm{T}}^{e}$ and $g_{\mathrm{T}}^{\nu_\ell}$ denote the tensor couplings to electrons and neutrinos of flavor $\ell$, respectively.  
The resulting contribution of this interaction to the E$\nu$ES cross section is given by \cite{DeRomeri:2022twg}
\begin{align}
\begin{split}
\biggl[\frac{d\sigma_{\nu_\ell}}{dT_{e}}\biggr]_{\mathrm{T}} =
&\frac{(g_\mathrm{T}^e g_\mathrm{T}^{\nu_\ell})^2}{2\pi (m_{\mathrm{T}}^2 + 2m_eT_{e})^2} m_e \Bigg[1 + 2\left(1-\frac{T_{e}}{E_\nu}\right) \\ & + \left(1-\frac{T_{e}}{E_\nu}\right)^2- \frac{m_e T_{e}}{E_\nu^2} \Bigg].
\end{split}
\label{eq:difcross_tensor}
\end{align}

\subsubsection{Lepton flavor-dependent $U(1)'$ models: $L_e-L_\mu$, $L_e-L_\tau$, $L_\mu-L_\tau$, and $L_e+2L_\mu+2L_\tau$}
The class of $U(1)'$ extensions considered here is purely leptonic in nature. These models constitute some of the simplest realizations of a new vector mediator $Z'$, as gauge anomaly cancellation can be achieved without introducing fermions beyond those already present in the SM \cite{He:1991qd}. Their phenomenology differs in several aspects from more traditional dark-photon scenarios. In particular, the associated gauge bosons do not couple to the charged weak bosons $W^\pm$ at tree level, and their interactions are restricted to the differences of lepton family numbers carried by charged leptons and neutrinos. They do not couple to baryon number.  
Because of these features, neutrino experiments provide particularly relevant probes of such models. A broad range of studies has explored their implications using stopped-pion neutrino sources \cite{AtzoriCorona:2022moj}, solar-neutrino measurements \cite{Coloma:2022umy, Gninenko:2020xys}, dark matter direct detection data \cite{DeRomeri:2024dbv, Demirci:2025qdp}, and collider searches \cite{Bauer:2018onh}.

\begin{table}[h]
    \caption{Individual vector charges of leptons and neutrinos for the lepton flavor-dependent $U(1)'$ models considered in this work.}
    \begin{ruledtabular}
	\begin{tabular}{c c c c c c c} 
%		\hline
%		\hline
		\textbf{Model} &  & $Q_{Z'}^{e/\nu_e}$ & &  $Q_{Z'}^{\mu/\nu_\mu}$ & & $Q_{Z'}^{\tau/\nu_\tau}$ \\
		\hline
%		universal & & $1$ & & $1$ & & $1$ \\
		$L_e-L_\mu$ & & $1$ & & $-1$ & & $0$ \\
		$L_e-L_\tau$ & & $1$ & & $0$ & & $-1$ \\
		$L_\mu-L_\tau$ & & $0$ & & $1$ & & $-1$\\
        $L_e+2L_\mu+2L_\tau$ & & $1$ & & $2$ & & $2$\\
%		\hline
%		\hline
	\end{tabular}
    \end{ruledtabular}
	\label{tab:modelcharge}
\end{table}
We summarize in Table~\ref{tab:modelcharge} the vector charges associated with each leptonic $U(1)'$ scenario. These charges determine the effective new-physics couplings entering E$\nu$ES. Accordingly, the $Z'$ contributions for the $L_e-L_\mu$, $L_e-L_\tau$, and $L_e+2L_\mu+2L_\tau$ models are obtained by directly inserting their respective charge assignments into Eq.~\eqref{eq:vec_chrg} and substituting them in Eq.~\eqref{eq:sm_nue}.

The situation differs for the $L_\mu-L_\tau$ model. In this case, electrons carry no charge under the new gauge group, implying that tree-level interactions with electrons are absent. As a result, the leading contribution to $\nu_\mu$--$e$ and $\nu_\tau$--$e$ scattering arises only at the one-loop level, while the effect on $\nu_e$--$e$ scattering first appears at two loops via $Z^0$–$Z'$ mixing and is therefore negligible.\footnote{The two-loop contribution to $\nu_e$--$e$ scattering is strongly suppressed and can safely be ignored.} The loop-induced corrections can be incorporated by performing the following replacement in Eq.~\eqref{eq:sm_nue} ~\cite{Altmannshofer:2019}:
\begin{align}
    g_V \;\longrightarrow\;
    g_V
    - 
    \frac{
        \sqrt{2}\,\alpha_{\mathrm{em}}\,
        g^e_{Z'}\, g^{\nu_\ell}_{Z'}\,
        (\delta_{\ell\mu}-\delta_{\ell\tau})
    }{
        \pi\, G_F \bigl(m_{Z'}^2 + 2 m_e T_e\bigr)
    }
    \,\epsilon_{\tau\mu}(|\vec{q}|),
    \label{eq:lmlt}
\end{align}
where the sign of the loop contribution differs for $\nu_\mu$ and $\nu_\tau$ scattering. Here $\alpha_{\mathrm{em}}$ is the electromagnetic fine-structure constant, and the loop factor $\epsilon_{\tau\mu}$ can be approximated as 
\begin{align}
    \epsilon_{\tau\mu}(|\vec{q}|)
    \simeq \frac{1}{6}\,
    \ln\!\left(\frac{m_\tau^2}{m_\mu^2}\right).
\end{align}

\section{Analysis Setup}\label{sec:stat}
In this section, we describe the statistical methodology employed to derive constraints on the mass-coupling parameters of light mediator models. Our analysis utilizes low-energy ER data collected by the LZ experiment during its first two science runs, WS2022~\cite{LZ:2022lsv} and WS2024~\cite{LZ:2024zvo}. The combination of these datasets yields a total integrated exposure of $4.2~\text{ton}\cdot\text{years}$~\cite{LZ:2025zpw}.

As an irreducible background in DD experiments, the SM E$\nu$ES contribution is theoretically well-characterized and exhibits an approximately flat spectrum with respect to recoil energy. Consequently, it is typically modeled as a known background component in standard dark matter searches. However, BSM physics scenarios can significantly enhance this rate, leading to potentially observable deviations from the SM expectation.
In this regard, we investigate the impact of light mediators on the neutrino–electron scattering spectrum and analyze their effects in the region of interest using the WS2022 and WS2024 datasets.

\subsection{Recoil Event Rates} \label{sec:ER}
The differential event rate is obtained by folding the E$\nu$ES cross section with the relevant solar neutrino flux. It can be expressed as  
\begin{align}
    \left[\frac{dR}{dT_{e}}\right]^{i}_{X}
    = 
    Z_{\mathrm{eff}}(T_e)\!
    \int_{E_{\nu}^{\mathrm{min}}}^{E_{\nu,i}^{\mathrm{max}}}\!
    dE_\nu\,
    \frac{d\Phi^{i}}{dE_\nu}\,
    \left[\frac{d\sigma}{dT_{e}}\right]^{\nu e}_{X},
\end{align}
where the lower integration limit is determined by kinematics,
\begin{align}
    E_{\nu}^{\mathrm{min}}
    = 
    \frac{1}{2}\left(T_{e}
    + \sqrt{T_{e}^{2} + 2m_{e} T_{e}}\right),
\end{align}
and $E_{\nu,i}^{\mathrm{max}}$ denotes the endpoint energy of each solar neutrino component. The index $i = pp,\,{}^{7}\mathrm{Be},\ldots$ labels the dominant solar neutrino fluxes, and in this work we restrict ourselves to the $pp$ and ${}^7\mathrm{Be}$ components, which are most relevant for low-energy neutrino–electron scattering. The label $X=\mathrm{SM},\,\phi,\,Z',\ldots$ corresponds to the considered interaction model.

At recoil energies of a few keV, the electron binding energies in xenon are no longer negligible, rendering the free-electron approximation insufficient \cite{Chen:2016eab}. To account for atomic ionization effects, we include the effective number of electrons that can participate in the scattering,
\begin{align}
    Z_{\mathrm{eff}}(T_e)
    = 
    \sum_{\alpha} n_\alpha\,\theta(T_e - B_\alpha),
\end{align}
where $n_\alpha$ and $B_\alpha$ denote, respectively, the electron multiplicity and binding energy of atomic shell $\alpha$ \cite{xraydata:2009}, and $\theta(x)$ is the Heaviside step function. This correction ensures that the cross section—originally derived under the assumption of free, stationary electrons—properly reflects the atomic structure of the target material \cite{Kouzakov:2017hbc, Hsieh:2019hug}.

Solar neutrinos undergo flavor oscillations while propagating from the solar core to the Earth. Consequently, the neutrino flux arriving at a detector consists of an admixture of $\nu_e$, $\nu_\mu$, and $\nu_\tau$. Therefore, the scattering cross section must be weighted by the corresponding flavor-transition probabilities, given by:
 \begin{align}
     \left[\frac{d\sigma}{dT_{e}}\right]^{\nu e}_X 
     = P_{ee}\left[\frac{d\sigma_{\nu_e}}{dT_{e}}\right]_{X}
     + \sum_{f=\mu,\tau} P_{ef}\left[\frac{d\sigma_{\nu_f}}{dT_{e}}\right]_{X},
 \end{align}
where the appearance probabilities for $\nu_e \to \nu_\mu$ and $\nu_e \to \nu_\tau$ take the form  
\begin{align}
    P_{e\mu} = (1-P_{ee})\cos^{2}\vartheta_{23}, \\
    P_{e\tau} = (1-P_{ee})\sin^{2}\vartheta_{23}.
\end{align}
The $\nu_e$ survival probability $P_{ee}$ is given by \cite{Maltoni:2015kca}
\begin{align}
\begin{split}
    P_{ee} =\;&
    \cos^{2}\vartheta_{13}\,\cos^{2}\vartheta_{13}^{\,m}
    \left[
        \frac{1}{2}
        + \frac{1}{2}
        \cos(2\vartheta_{12}^{\,m})\cos(2\vartheta_{12})
    \right]
    \\
    &+ 
    \sin^{2}\vartheta_{13}\,\sin^{2}\vartheta_{13}^{\,m},
\end{split}
\label{Pee}
\end{align}
where the superscript $m$ denotes quantities modified by solar matter effects. The survival and transition probabilities depend on the mixing angles $\vartheta_{12}$, $\vartheta_{13}$, and $\vartheta_{23}$.

In our analysis, we incorporate the day–night asymmetry arising from Earth matter effects when evaluating these probabilities. For numerical inputs, we adopt the normal-ordering oscillation parameters from the recent NuFit~6.0 global fit \cite{Esteban:2024eli}, excluding Super-Kamiokande atmospheric data, as recommended for solar neutrino studies.

The predicted number of E$\nu$ES events is obtained from
\begin{widetext}
\begin{align}
\begin{split}
	R^k_X=&\varepsilon N_{t}\int_{T_e^k}^{T_e^{k+1}}dT_e \text{ } \mathcal{A}(T_e) \int_{0}^{T_e^{'\text{max}}}dT_e' \text{ } \mathcal{R}(T_e,T_e') %\\ &\times 
    \sum_{i=pp,^7\text{Be}} \left[\frac{dR}{dT'_{e}}\right]^i_X,
\end{split}
\end{align}
\end{widetext}
where the factor $N_{t}$ denotes the number of target nuclei per unit mass of the detector material and the maximum recoil energy satisfies $T_e^{'\text{max}} = \frac{2E_\nu^2}{2E_\nu + m_e}$.
The reconstructed and true electron recoil energies are denoted by $T_e$ and $T_e'$, respectively. The detection efficiency is denoted by $\mathcal{A}(T_e)$, where we use the 2D ROI efficiencies of the LZ datasets from Ref.~\cite{LZ:2025zpw}. Meanwhile, for the smearing functions $\mathcal{R}(T_e,T_e')$, we use the normalized Gaussian form with $\sigma = 0.2325 \sqrt{T_e}$ energy resolution \cite{Pereira:2023rte}. These two factors should be taken into account to simulate the ER signal accurately. 
The predicted event-rate also includes the exposure factor $\varepsilon$, where the LZ WS2022 dataset has $0.9\text{ton} \cdot \text{year}$ and LZ WS2024 dataset has $3.3 \text{ton} \cdot \text{year}$.

\subsection{$\chi^2$-Function}
To derive constraints on the new physics parameter(s) of interest, denoted by $\mathcal{S}$, we employ a Poisson-based $\chi^2$ test statistic~\cite{Baker:1983tu, Fogli:2002pt}, which is defined as the sum of a modified log-likelihood ratio over all energy bins $k$ and $\chi^2$ pull terms that constrain the nuisance parameters $\alpha_i$ and $\beta_j$:
\begin{widetext}
\begin{align}
\begin{split}
\chi^2(\mathcal{S}, \alpha_i, \beta_j) = &2  \sum_{k=1}^{20} \biggl[  R_\text{exp}^{k}(\mathcal{S}; \alpha,\beta) - R_{obs}^{k} + R_{obs}^{k} \ln \Bigg( \frac{R_{obs}^{k} }{R_\text{exp}^{k}(\mathcal{S}; \alpha,\beta) }\Bigg)  \biggr] +\sum_{i}\left(\frac{\alpha_i}{\sigma_{\alpha_i}}\right)^2 +  \sum_{j}\left(\frac{\beta_j}{\sigma_{\beta_j}}\right)^2,
\end{split} \label{eq:chi2}
\end{align} 
\end{widetext}
where $R_{\text{obs}}^k$ and $R_{\text{exp}}^k$ represent the observed and expected event rates in the $k^{\text{th}}$ energy bin, respectively. The expected event rate, $R_{\text{exp}}^k$, consists of the SM contribution $R^k_{\text{SM}}$, plus the New Physics (NP) contribution $R^k_{\text{NP}}$ (from sources like $\phi, Z', \cdots$), and various background components $R_{\text{Bkg},j}^k$. Explicitly, the expected rate is given by
\begin{align}
\begin{split}
R_{\text{exp}}^k(\mathcal{S};\alpha_i, \beta_j)=& (1+\alpha_i)(R^k_{\text{SM}}+R^k_{\text{NP}}(\mathcal{S})) \\ & + (1+\beta_j)R_{\text{Bkg},j}^k
\end{split} \label{eq:Rexp}
\end{align} 
where $\alpha_i$ and $\beta_j$ denote the nuisance parameters associated with the uncertainties in the solar neutrino flux components and the various background contributions, respectively. The minimum value is found by minimizing the function with respect to the nuisance parameters $\alpha_i$ and $\beta_j$:
\begin{align}
    \chi^2_{\min}(\mathcal{S}) = \min_{\alpha_i, \beta_j} \left[ \chi^2(\mathcal{S}, \alpha_i, \beta_j) \right]
\end{align}

The solar neutrino flux uncertainty is denoted by $\sigma_{\alpha}$, with values of $0.6\%$ and $3\%$ adopted for the $pp$ and $^{7}\text{Be}$ fluxes, respectively~\cite{Baxter:2021pqo}. The uncertainties of the background components, $\sigma_{\beta_j}$, are taken from Ref.~\cite{LZ:2025zpw}. The SM prediction supplemented with possible new-physics contributions is evaluated using the solar neutrino fluxes from the Bahcall spectrum \cite{Bahcall:1989ks}, normalized according to the B16-GS98 standard solar model \cite{Vinyoles:2016djt}. With this input, the total $\nu e$ scattering event rates—obtained by summing over all recoil-energy bins—are found to be consistent with the corresponding rates reported by the LZ Collaboration.

Based on this methodology, we calculate the $90\%$ C.L. upper limits using a two-parameter $\chi^2$ analysis~\eqref{eq:chi2} of the LZ WS2022 and WS2024 low-energy electron recoil datasets. Furthermore, we perform a combined analysis of both datasets by accounting for the correlated neutrino flux uncertainties.

%%%%%%%%%%%%%%%%%%%%%
%%%%%%%%%%%%%%%%%%%%%
\begin{figure*}[htb]
	\centering
	\includegraphics[scale=0.40]{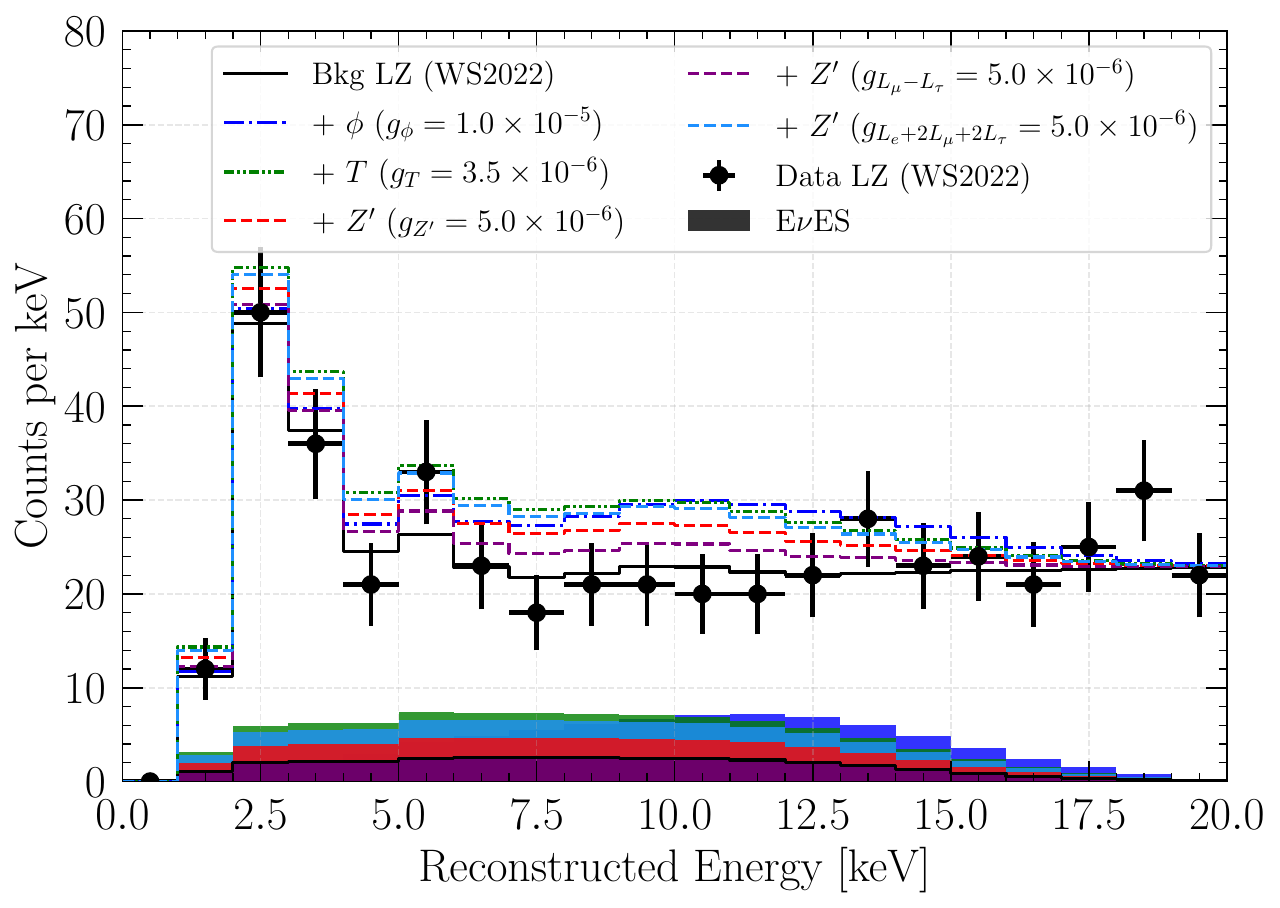}
	\includegraphics[scale=0.40]{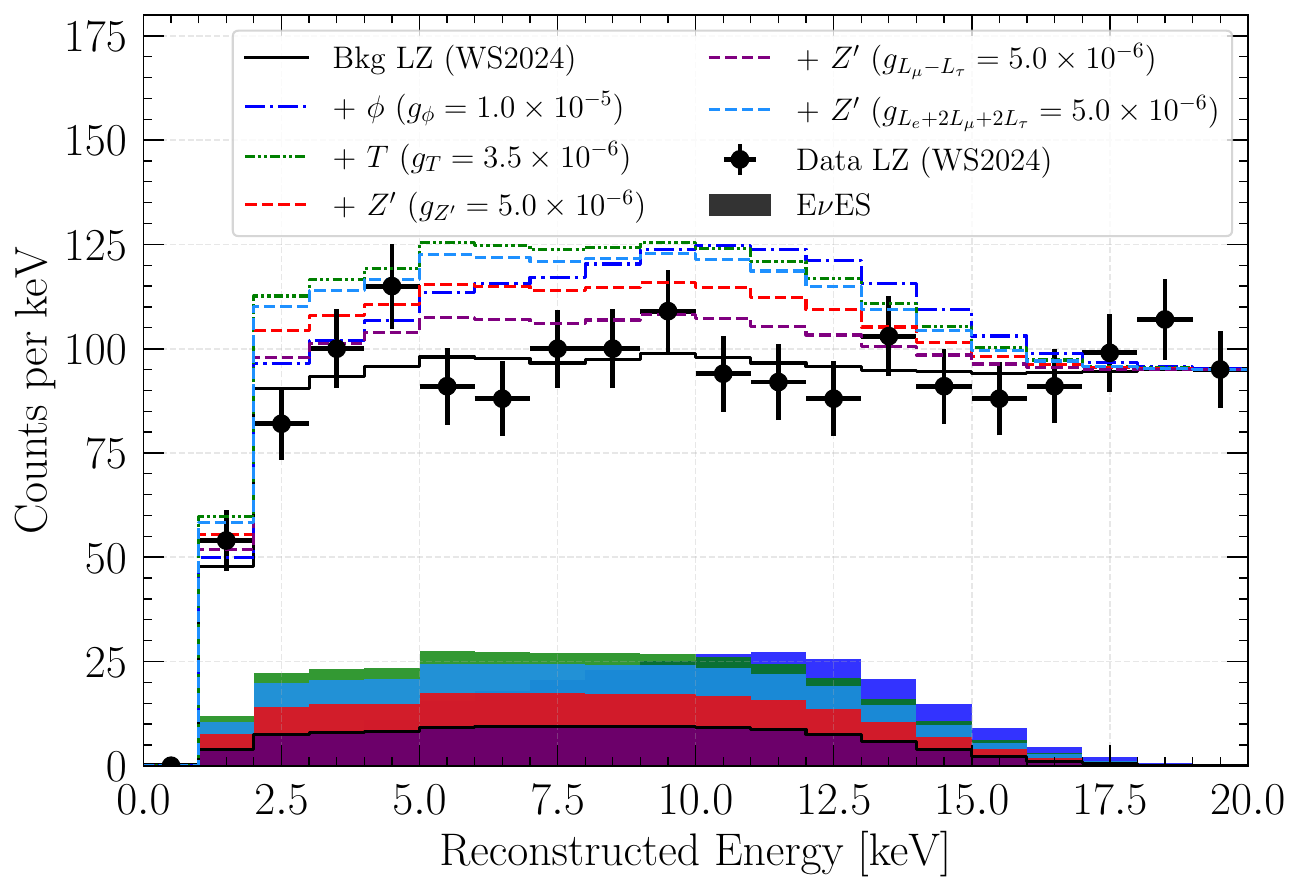}
	\\
	\hspace{7mm} (a) \hspace{82mm} (b)
	\caption{ Electron recoil energy distributions for light mediator models and LZ data (black points with error bars) in the (a) WS2022 and (b) WS2024 datasets. Filled histograms denote individual E$\nu$ES signal contributions; outlined histograms of the same color show their respective contributions to the total background. The predicted events are evaluated for a benchmark mass $m_X=1$ MeV and various couplings as shown in the legend.
    }
	\label{fig:cevns_bsm}
\end{figure*}
%%%%%%%%%%%%%%%%%%%%%
%%%%%%%%%%%%%%%%%%%%%
\subsection{Constraints from Other Experiments}\label{sec:exist_constraints}
For comparison, we overlay our results with the currently available bounds obtained from a comprehensive set of experimental probes. These include limits from solar neutrino experiments (BOREXINO \cite{Coloma:2022umy,Gninenko:2020xys}), nuclear reactor experiments (GEMMA ($\bar{\nu}_e e$ scattering) \cite{Lindner:2020kko}, TEXONO ($\overline{\nu}_e e$ scattering) \cite{Bilmis:2015lja}, DRESDEN-II (CE$\nu$NS)\cite{Coloma:2022avw}, CONNIE (CE$\nu$NS) \cite{CONNIE:2019xid}, CONUS ($\bar{\nu}_e e$ scattering) \cite{CONUS:2021dwh}, and CONUS+ (CE$\nu$NS+$\bar{\nu}_e e$ scattering) \cite{Chattaraj:2025}), the stopped-pion experiment (COHERENT (combined CsI+LAr analysis) \cite{DeRomeri:2022twg}), other neutrino--electron scattering experiments (CHARM-II ($\overline{\nu}_\mu e$ scattering) \cite{Bilmis:2015lja}, and LSND ($\nu_e e$ scattering) \cite{Bilmis:2015lja}), as well as leading dark matter direct detection experiments (PANDAX-4T (Run1) ($\nu e$ scattering) \cite{Demirci:2025qdp}, XENONnT ($\nu e$ scattering) \cite{Melas:2023olz,A:2022acy}, and CDEX-10 (CE$\nu$NS channel) \cite{Demirci:2023tui}). 

Furthermore, we include several collider limits obtained from experiments such as NA64 \cite{NA64:2022yly}, Mainz \cite{A1:2011yso}, KLOE \cite{ALICE:2012aqc}, BaBar \cite{BaBar:2014zli}, PHENIX \cite{PHENIX:2014duq}, and NA48/2 \cite{NA482:2015wmo}. We also incorporate constraints from neutrino oscillation data \cite{Coloma:2022umy,Coloma:2020gfv} and the favored parameter space associated with the muon anomalous magnetic moment $(g-2)_\mu$ \cite{Muong-2:2023cdq}.

In addition, we present complementary astrophysical and cosmological bounds. These limits are derived from Big Bang Nucleosynthesis (BBN) \cite{Blinov:2019gcj,Huang:2018}, supernova cooling observations (SN1987A) \cite{Heurtier:2017,Croon:2020lrf}, constraints on the Hubble constant ($H_0$) \cite{Escudero:2019gzq}, and stellar cooling processes \cite{LiXu:2023}. It should be emphasized that the precise predictions of these bounds can depend sensitively on the underlying new physics model and the assumed thermal history of the early Universe. The full details of our results are presented in the subsequent section.

\section{Results and Discussion}\label{sec:resdis}
In this section, we report the numerical findings derived from our statistical analysis. The low-energy ER data from the LZ Experiment are confronted with the predicted event rates associated with the light-mediator scenarios considered in this work, as shown in Fig.~\ref{fig:cevns_bsm}(a) for WS2022 and Fig.~\ref{fig:cevns_bsm}(b) for WS2024. All spectra are normalized in units of $\mathrm{keV}^{-1}$. We compute the E$\nu$ES signal, for which the dominant contributions originate from the $pp$ and $^7$Be solar neutrino fluxes. The corresponding predicted energy spectrum has been subtracted from the total background. For illustrative purposes, we fix the mediator mass to $m = 1\,\mathrm{MeV}$ and choose the benchmark coupling values as $g_\phi = 1.0\times10^{-5}$, $g_T = 3.5\times10^{-6}$, and $g_{Z'} = g_{L_\mu-L_\tau} = g_{L_e+2L_\mu+2L_\tau} = 5.0\times10^{-6}$. The individual contribution of each model is evaluated by embedding the corresponding interaction into the E$\nu$ES signal.

%%%%%%%%%%%%%%%%%%%%%
%%%%%%%%%%%%%%%%%%%%%
\begin{figure*}[htb]
	\centering
    	\includegraphics[scale=0.50]{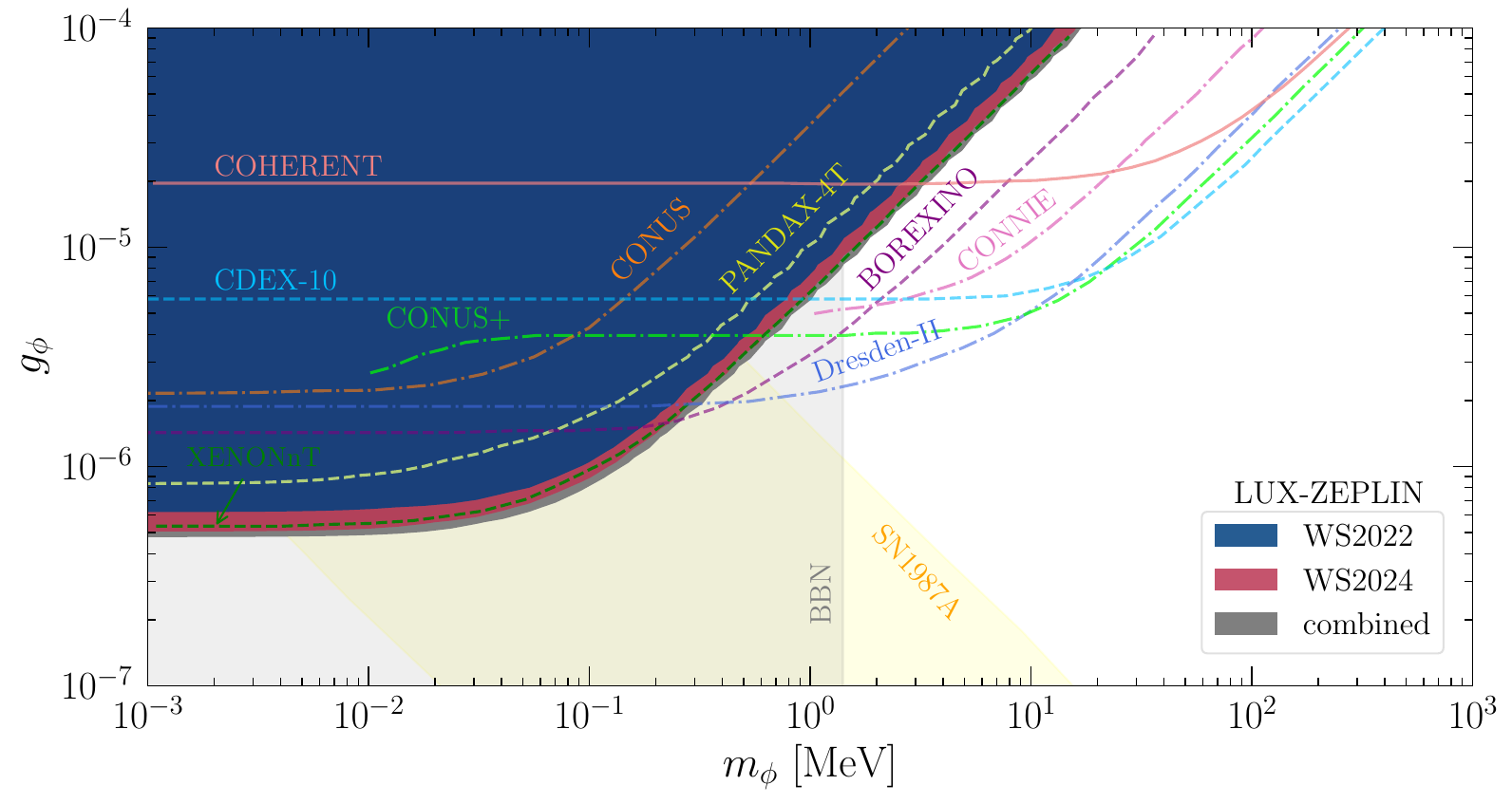}
	\\
	%\vspace{-2mm}
	\caption{The 90\% C.L. upper limits on the mass-coupling plane of the universal light scalar mediator from the LZ WS2022 and LZ WS2024 datasets, including a combined analysis. These are compared with existing limits obtained from previous works for COHERENT  \cite{DeRomeri:2022twg}, CONUS \cite{CONUS:2021dwh}, CONUS+ \cite{Chattaraj:2025}, %DUNE \cite{Melas:2023olz}, 
    CONNIE \cite{CONNIE:2019xid}, DRESDEN-II (Fef) \cite{Coloma:2022avw}, BOREXINO \cite{Coloma:2022umy}, PANDAX-4T \cite{Demirci:2025qdp}, XENONnT \cite{A:2022acy} and CDEX-10 \cite{Demirci:2023tui}. Astrophysical limits from BBN \cite{Blinov:2019gcj} and SN1987A \cite{Heurtier:2017} are also shown.
	}
	\label{fig:analysis_s}
\end{figure*}
%%%%%%%%%%%%%%%%%%%%%
%%%%%%%%%%%%%%%%%%%%%
From Fig.~\ref{fig:cevns_bsm}, it is evident that the inclusion of light mediator contributions can sizably distort the reconstructed electron recoil energy spectra with respect to the SM background expectation. In both WS2022 [Fig.~\ref{fig:cevns_bsm}(a)] and WS2024 [Fig.~\ref{fig:cevns_bsm}(b)] datasets, the excess induced by light mediators primarily appears in the low-energy region below $\sim 17$~keV, where the solar neutrino flux dominates, and the detector sensitivity is highest. Among the vector mediator scenarios, the $L_e+2L_\mu+2L_\tau$ model yields the largest enhancement in the event rate for the chosen benchmark couplings. We note that the corresponding spectra of the $L_e-L_\mu$ and $L_e-L_\tau$ models are not shown since their contribution is nearly identical to that of the universal $Z'$ mediator, rendering the curves visually indistinguishable. This degeneracy originates from the similar effective coupling structures governing E$\nu$ES in these three scenarios.

A clear improvement in statistical precision is observed when comparing WS2024 with WS2022, as reflected by the reduced error bars and the higher overall event statistics. This enhanced sensitivity significantly strengthens the capability of the WS2024 dataset to probe light-mediator-induced deviations in the low-energy regime. Overall, these results demonstrate that light mediators provide an efficient and robust mechanism to amplify low-energy electron recoil signals, thereby substantially extending the new-physics reach of liquid-xenon direct detection experiments.

%%%%%%%%%%%%%%%%%%%%%
%%%%%%%%%%%%%%%%%%%%%
\begin{figure*}[htb]
	\centering
	\includegraphics[scale=0.50]{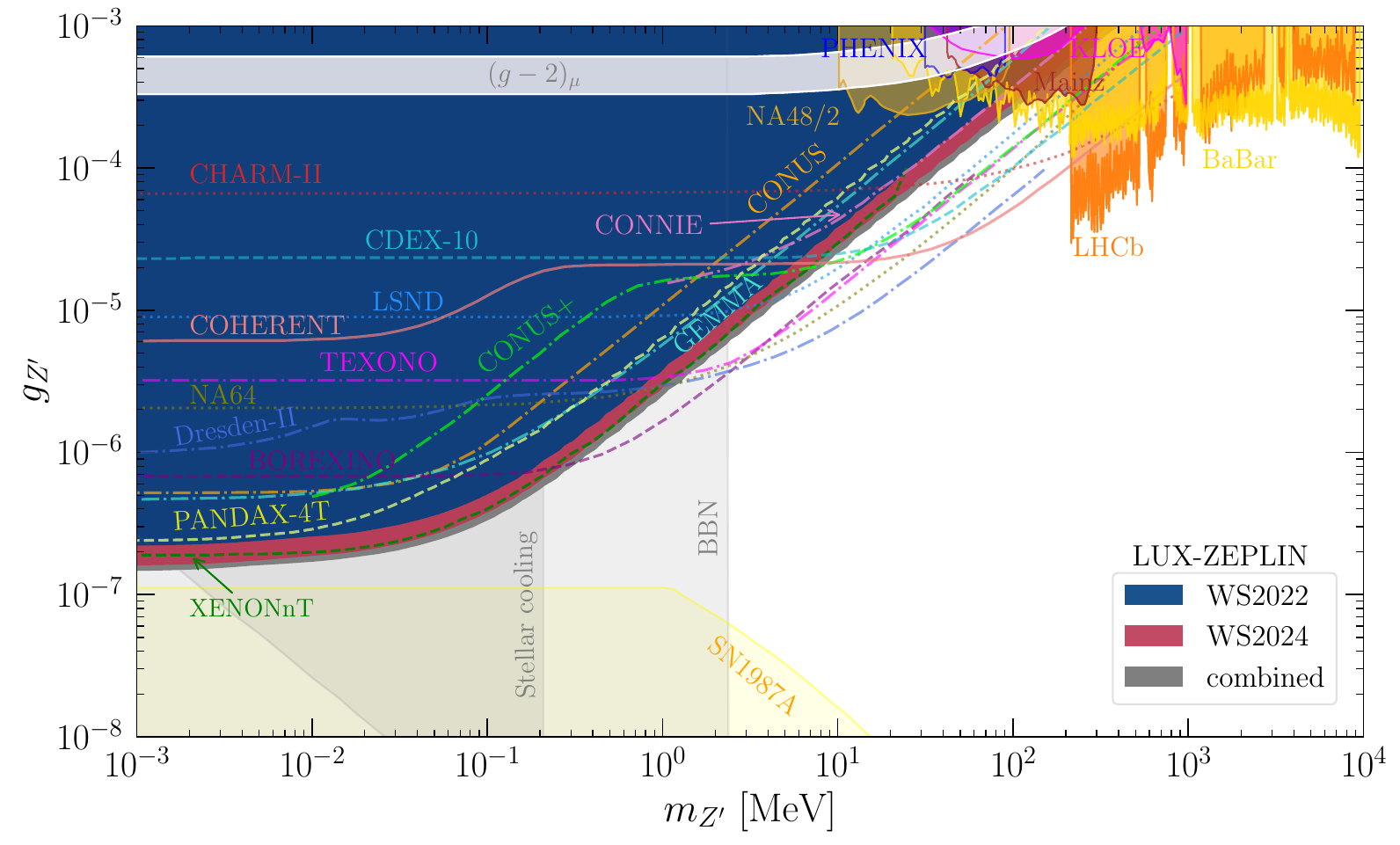}
	\\
	%\vspace{-2mm}
	\caption{The 90\% C.L. upper limits on the mass-coupling plane of the universal light vector mediator from the LZ WS2022 and LZ WS2024 datasets, including a combined analysis. These are compared with existing limits obtained from previous works for BOREXINO \cite{Coloma:2022umy}, COHERENT  \cite{DeRomeri:2022twg}, CONUS \cite{CONUS:2021dwh}, CONNIE \cite{CONNIE:2019xid}, CONUS+ \cite{Chattaraj:2025}, TEXONO \cite{Bilmis:2015lja}, DRESDEN-II (Fef) \cite{Coloma:2022avw},  GEMMA \cite{Lindner:2020kko}, CHARM-II \cite{Bilmis:2015lja}, LSND \cite{Bilmis:2015lja}, NA64 \cite{NA64:2022yly}, PANDAX-4T \cite{Demirci:2025qdp}, XENONnT \cite{A:2022acy} and CDEX-10 \cite{Demirci:2023tui}. The limits from colliders (Mainz \cite{A1:2011yso}, KLOE \cite{ALICE:2012aqc}, BaBar \cite{BaBar:2014zli}, PHENIX \cite{PHENIX:2014duq}, NA48/2 \cite{NA482:2015wmo}) and astrophysical limits (BBN ($\Delta N_{\text{eff}}\simeq 1)$ \cite{Huang:2018}, stellar cooling \cite{LiXu:2023}, SN1987A \cite{Croon:2020lrf}) as well as the allowed bound of $(g-2)_\mu$ \cite{Muong-2:2023cdq} are also shown.}
	\label{fig:analysis_v}
\end{figure*}
%%%%%%%%%%%%%%%%%%%%%
%%%%%%%%%%%%%%%%%%%%%
We present our results in Fig.~\ref{fig:analysis_s} in the coupling–mass plane for the universal scalar mediator at 90\% C.L. (2 d.o.f.).
The exclusion limits derived from the WS2022 dataset are shown in blue, those from WS2024 in red, while the grey region corresponds to the limit obtained from the combined analysis of the two datasets. For mediator masses $m_\phi \leq 1$ keV, the resulting upper limits on the coupling strength are $g_\phi = 6.17\times 10^{-7}$ for WS2022, $5.03\times 10^{-7}$ for WS2024, and $4.76\times 10^{-7}$ for the combined analysis. As expected, owing to its larger exposure, the WS2024 dataset alone yields an approximately 1.2 times stronger constraint compared to WS2022. Moreover, the combined analysis further improves the sensitivity by coherently exploiting the statistical power of both datasets, leading to the strongest bound obtained in this work and demonstrating the complementarity of the two running periods. Furthermore, the limits of previous studies discussed in Sec.~\ref{sec:exist_constraints} are also outlined for comparison. We can see that both datasets provide more stringent limits than those obtained from BOREXINO, CONNIE (95\% C.L.), CDEX-10, COHERENT, CONUS+, and Dresden-II in the low-mass region, while they cover the limit derived from CONUS and PANDAX-4T.
Meanwhile, although the exclusion limits obtained from the LZ datasets are of the same order of magnitude as those from XENONnT, the WS2024 dataset yields slightly more stringent constraints, particularly in the low-mass region. It is clearly seen that the combination limit is even more stringent, demonstrating the superior sensitivity achieved with the increased exposure. 
Additionally, this analysis offers valuable constraints that can reach some parts of the existing bounds from BBN and SN1987A.

%%%%%%%%%%%%%%%%%%%%%
%%%%%%%%%%%%%%%%%%%%%
\begin{figure*}[htb]
	\centering
    	\includegraphics[scale=0.50]{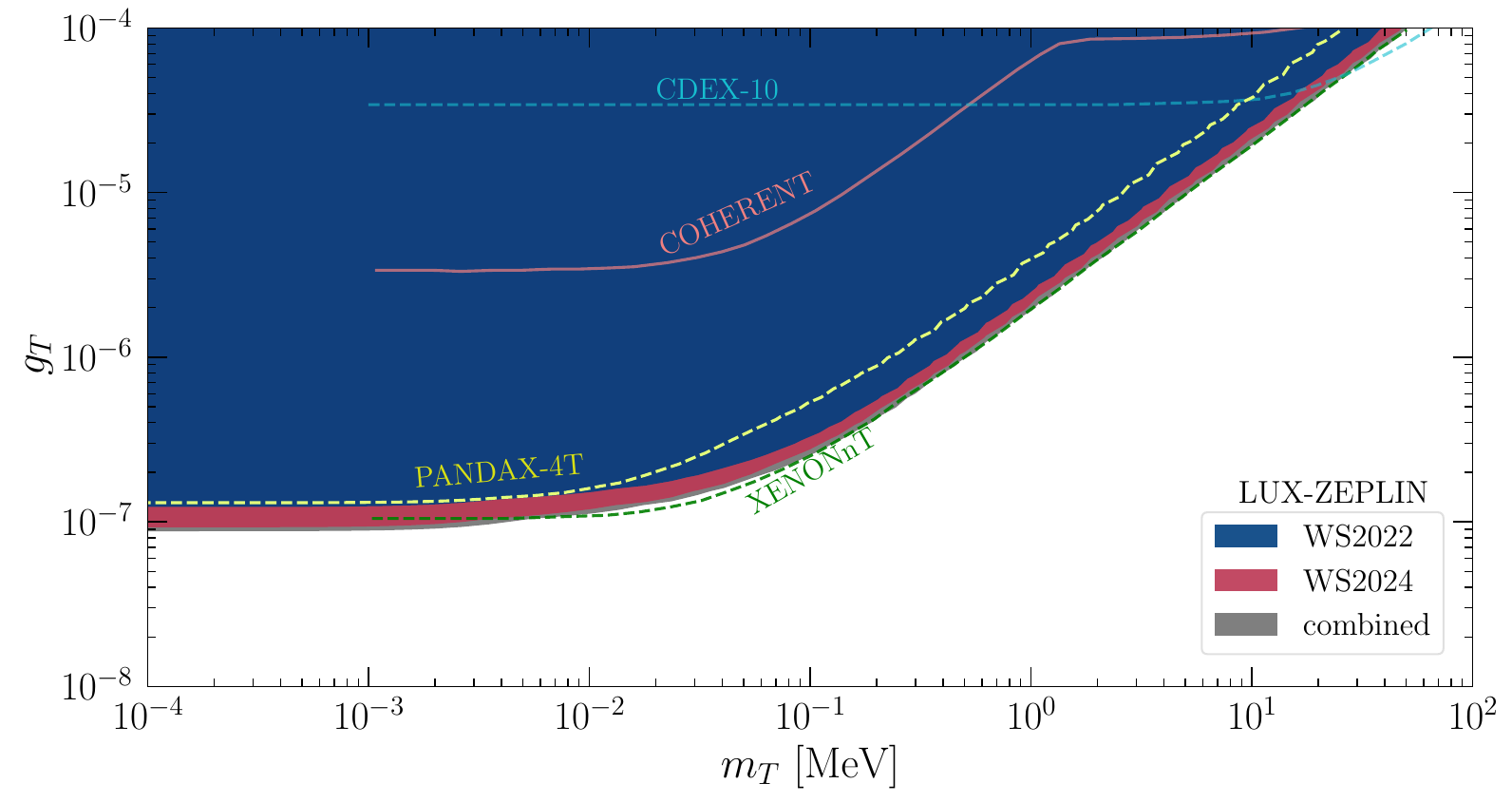}
	\\
	%\vspace{-2mm}
	\caption{The 90\% C.L. upper limits on the mass-coupling plane of the universal light tensor mediator from the LZ WS2022 and LZ WS2024 datasets, including a combined analysis. These are compared with existing limits obtained from previous works for COHERENT \cite{DeRomeri:2022twg}, PANDAX-4T \cite{Demirci:2025qdp}, XENONnT \cite{A:2022acy} and CDEX-10 \cite{Demirci:2023tui}. 
	}
	\label{fig:analysis_t}
\end{figure*}
%%%%%%%%%%%%%%%%%%%%%
%%%%%%%%%%%%%%%%%%%%%
%%%%%%%%%%%%%%%%%%%%%
%%%%%%%%%%%%%%%%%%%%%
\begin{figure*}[htb]
	\centering
	\includegraphics[scale=0.50]{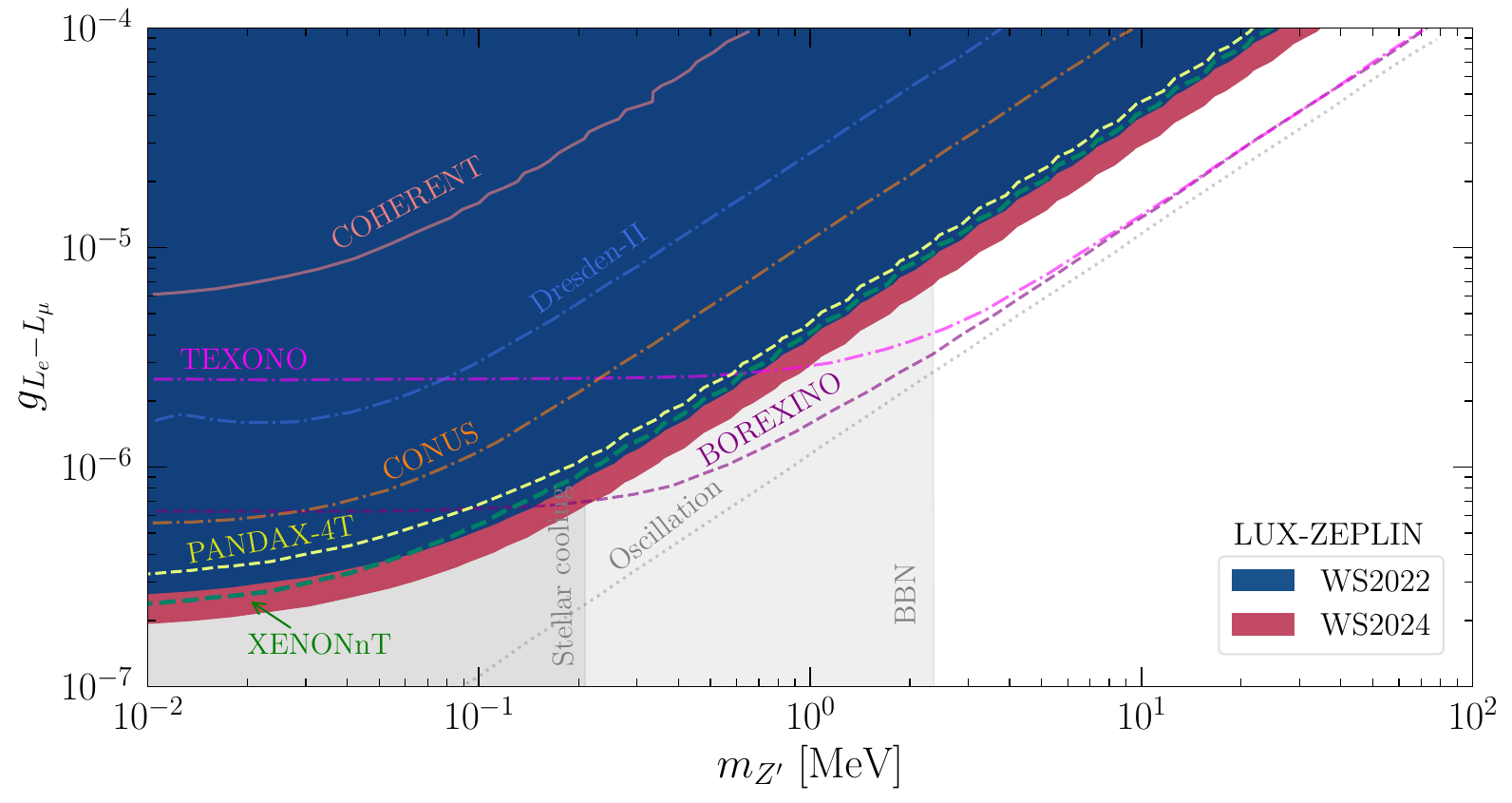}
	\\
	\caption{The 90\% C.L. upper limits on the mass-coupling plane of the $L_e-L_\mu$ model from the WS2022 and WS2024 datasets of LZ Experiment. These are compared with existing limits obtained from previous works for COHERENT \cite{Coloma:2022avw, Coloma:2022umy}, CONUS \cite{Coloma:2022umy}, TEXONO \cite{Coloma:2022umy, Coloma:2020gfv}, DRESDEN-II \cite{Coloma:2022avw, Coloma:2022umy}, BOREXINO \cite{Coloma:2022umy}, and, derived in this work, PANDAX-4T and XENONnT.
    The limits obtained from a global fit to oscillation data \cite{Coloma:2022umy, Coloma:2020gfv},  BBN ($\Delta N_{\text{eff}}\simeq 1)$ \cite{Huang:2018} and stellar cooling \cite{LiXu:2023} are also shown.
	}
	\label{fig:analysis_lelm}
\end{figure*}
%%%%%%%%%%%%%%%%%%%%%
%%%%%%%%%%%%%%%%%%%%%

In Fig.~\ref{fig:analysis_v}, we present the upper limits in the coupling–mass plane for the universal vector mediator obtained from the datasets analyzed in this work. In the low-mass regime, $m_{Z'} \leq 1,\mathrm{keV}$, the corresponding upper bounds on the coupling strength are $g_{Z'} = 2.20\times10^{-7}$ for WS2022, $1.61\times10^{-7}$ for WS2024, and $1.46\times10^{-7}$ for the combined analysis. Owing to its larger exposure, the WS2024 dataset yields an improvement in sensitivity by approximately a factor of 1.4 with respect to WS2022 in the low-mass region.
We also superimpose the existing limits for comparison. In the low-mass region, our results set the most stringent constraints to date, surpassing all previously reported bounds. In particular, for mediator masses below $m_{Z'} \lesssim 6\,\mathrm{MeV}$, our obtained results fully overtake the COHERENT constraint, resulting in an improvement of approximately one order of magnitude. We also observe that more stringent constraints are obtained than those derived from nuclear reactor experiments (CONNIE (95\% C.L.), TEXONO, Dresden-II, and CONUS+), the DD experiment (CDEX-10, PANDAX-4T), and other electron-neutrino scattering experiments (CHARM-II, LSND, and NA64). The obtained limits from the WS2022 and WS2024 datasets entirely cover the GEMMA and CONUS limits, and provide a few times improvement over the BOREXINO limit as $m_{Z'}\leq 0.3$ MeV. Moreover, the limit from WS2024 is of the same order of magnitude as the XENONnT, where the obtained result provides a modest improvement for $m_{Z'}\lesssim 0.01$ MeV. With the current datasets, the allowed bound of $(g-2)_\mu$ is completely covered for this model. 
Additionally, the obtained limits are complementary to the BBN, stellar cooling, and SN1987A in the low-mass region and collider limits in the high-mass region.

\begin{figure*}[htb]
	\centering
	\includegraphics[scale=0.50]{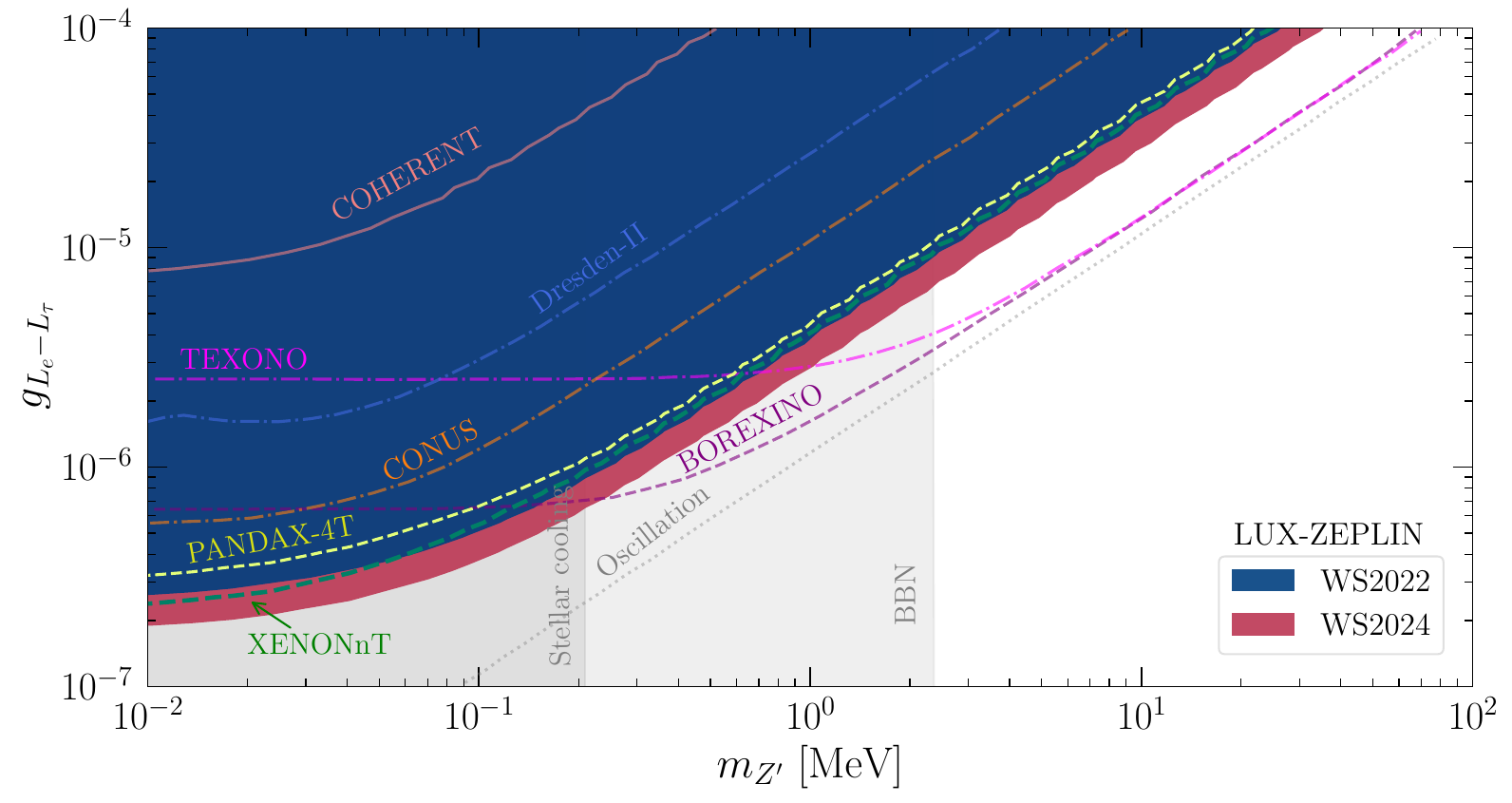}
	\\
	\caption{The 90\% C.L. upper limits on the mass-coupling plane of the $L_e-L_\tau$ model from the WS2022 and WS2024 datasets of LZ Experiment. These are compared with existing limits of COHERENT \cite{Coloma:2022avw, Coloma:2022umy}, CONUS \cite{Coloma:2022umy}, TEXONO \cite{Coloma:2022umy, Coloma:2020gfv}, DRESDEN-II \cite{Coloma:2022avw, Coloma:2022umy}, BOREXINO \cite{Coloma:2022umy}, and, derived in this work, PANDAX-4T and XENONnT.
    The limits obtained from a global fit to oscillation data \cite{Coloma:2022umy, Coloma:2020gfv},  BBN ($\Delta N_{\text{eff}}\simeq 1)$ \cite{Huang:2018} and stellar cooling \cite{LiXu:2023} are also shown.
	}
	\label{fig:analysis_lelt}
\end{figure*}
%%%%%%%%%%%%%%%%%%%%%
%%%%%%%%%%%%%%%%%%%%%
\begin{figure*}[htb]
	\centering
	\includegraphics[scale=0.50]{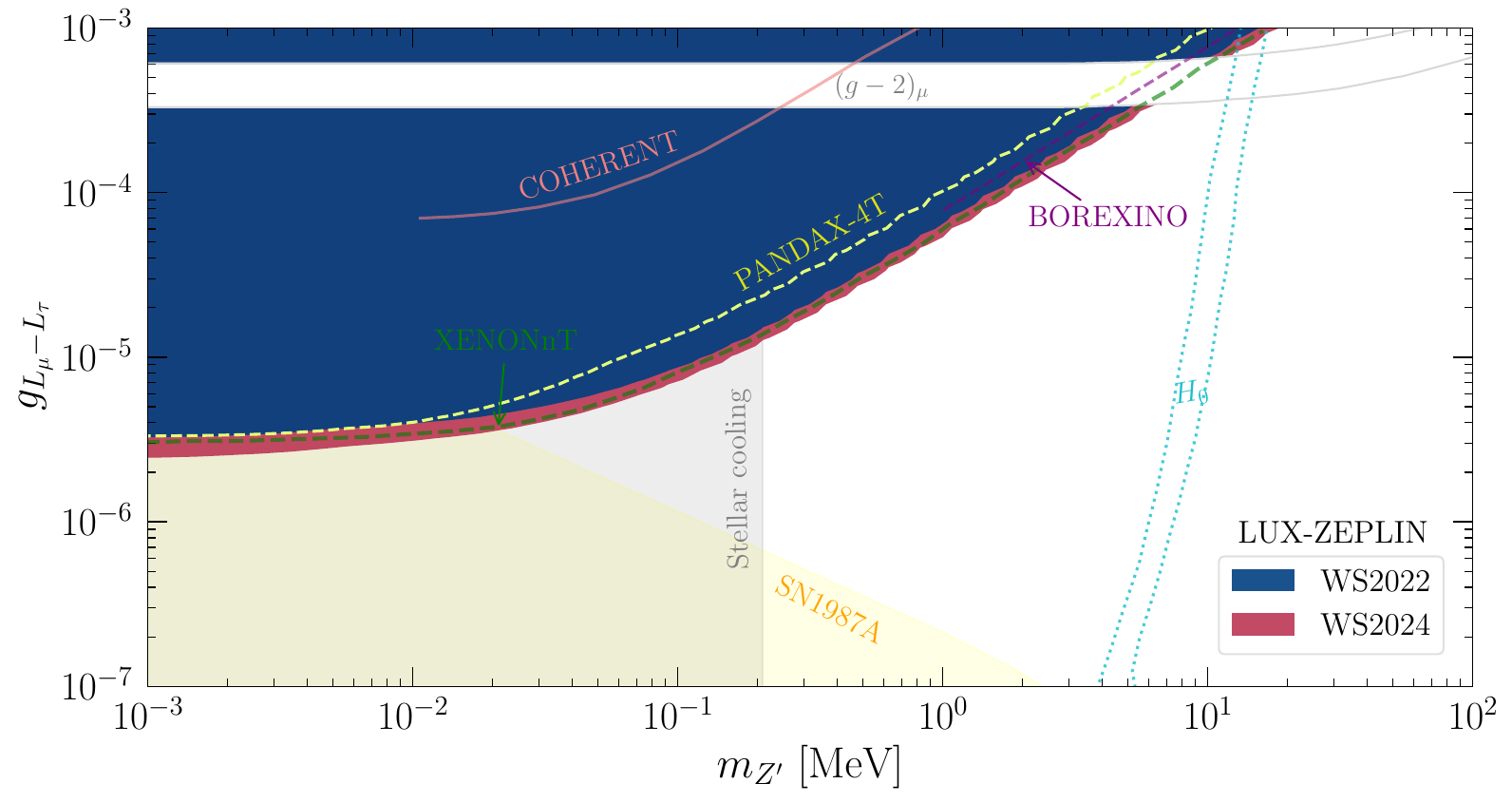}
	\\
	\caption{The 90\% C.L. upper limits on the mass-coupling plane of the $L_\mu-L_\tau$ model from the WS2022 and WS2024 datasets of LZ Experiment. These are compared with the existing limits of COHERENT and BOREXINO \cite{Melas:2023olz, Gninenko:2020xys}, PANDAX-4T \cite{Demirci:2025qdp}, and XENONnT \cite{Demirci:2025qdp}. The limits from SN1987A \cite{Croon:2020lrf}, stellar cooling \cite{LiXu:2023} and cosmology (the reported Hubble tension $H_0$) \cite{Escudero:2019gzq} as well as the allowed bound of $(g-2)_\mu$ \cite{Muong-2:2023cdq} are also shown.
	}
	\label{fig:analysis_lmlt}
\end{figure*}
%%%%%%%%%%%%%%%%%%%%%
%%%%%%%%%%%%%%%%%%%%%
\begin{figure*}[htb]
	\centering
	\includegraphics[scale=0.50]{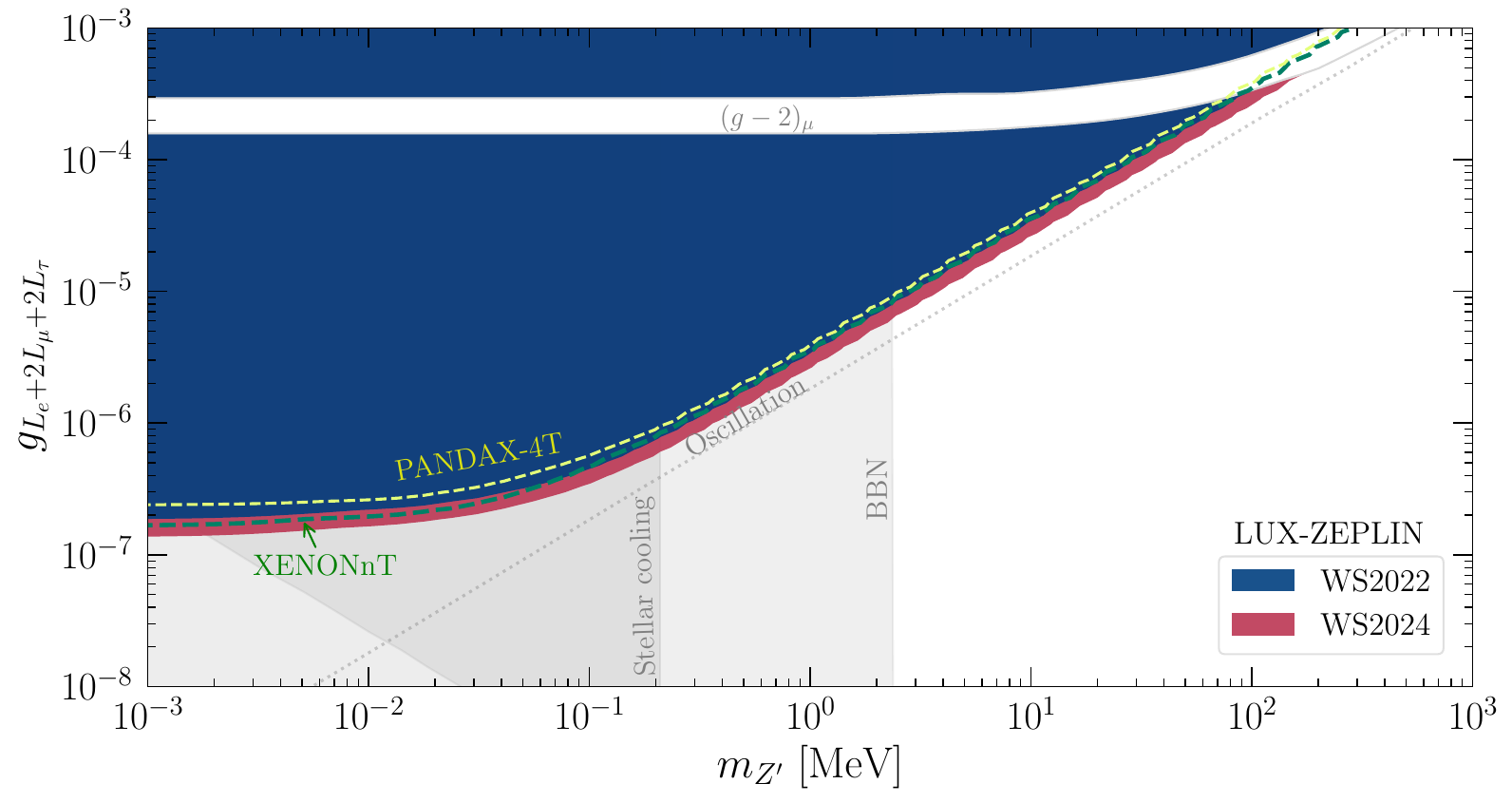}
	\\
	\caption{The 90\% C.L. upper limits on the mass-coupling plane of the $L_e+2L_\mu+2L_\tau$ model from the WS2022 and WS2024 datasets of LZ Experiment. The figure also displays the PANDAX-4T and XENONnT limits derived in this study, alongside existing bounds from global oscillation fits \cite{Coloma:2022umy, Coloma:2020gfv}, BBN ($\Delta N_{\text{eff}}\simeq 1)$ \cite{Huang:2018}, stellar cooling \cite{LiXu:2023}, and the $(g-2)_\mu$ allowed region \cite{Muong-2:2023cdq}.	}
	\label{fig:analysis_le2lmt}
\end{figure*}
%%%%%%%%%%%%%%%%%%%%%
%%%%%%%%%%%%%%%%%%%%%
Next, in Fig.~\ref{fig:analysis_t}, we present the exclusion limits for the universal light tensor mediator model. In the low-mass region with $m_{\mathrm{T}} \leq 1$ keV, the upper bounds on the coupling are found to be $g_{\mathrm{T}} = 1.23\times10^{-7}$ for WS2022, $9.23\times10^{-8}$ for WS2024, and $8.79\times10^{-8}$ for the combined analysis. The WS2024 dataset provides an improvement of a few times relative to WS2022, while the combined analysis yields the most stringent bound.
The existing limits from previous studies are also superimposed for comparison. Our results substantially improve upon the constraints from COHERENT and CDEX-10, exceeding them by roughly one and two orders of magnitude, respectively. The obtained limits are comparable to those from PANDAX-4T,
%—whose sensitivity remains stronger over the full mass range—
while the WS2024 result starts to exceed the XENONnT sensitivity for mediator masses below $3$ keV.

We now discuss the 90\% C.L. exclusion limits for the $L_e-L_\mu$, $L_e-L_\tau$, $L_\mu-L_\tau$, and $L_e+2L_\mu+2L_\tau$ light mediator models, presented in Figs.~\ref{fig:analysis_lelm}, \ref{fig:analysis_lelt}, \ref{fig:analysis_lmlt}, and \ref{fig:analysis_le2lmt}, respectively. For the $U(1)_{L_e+2L\mu+2L_\tau}$ case, we additionally derive limits using the latest electronic-recoil datasets from XENONnT \cite{XENON:2022ltv} and PandaX-4T \cite{PandaX:2024cic}, enabling a direct comparison across all currently available LXe experiments.

For the $L_e-L_\mu$ model, in the region of $m_{Z'}\lesssim 10^{-3}$ MeV, we obtain upper limits of $g_{L_e-L_\mu} = 2.28\times10^{-7}$ from WS2022 and $1.66\times10^{-7}$ from WS2024, demonstrating that the latter improves the sensitivity by a factor of approximately 1.37.
Similarly, for the $L_e-L_\tau$ model, the bounds reach $2.26\times10^{-7}$ for WS2022 and $1.61\times10^{-7}$ for WS2024, corresponding to an improvement of about a factor of 1.40.
Comparing our results with previously reported bounds, we find that the limits from both the $L_e-L_\mu$ and $L_e-L_\tau$ models entirely cover the COHERENT, Dresden-II, and CONUS constraints. %by more than an order of magnitude. 
The TEXONO bound is surpassed for mediator masses below $m_{Z'} \leq 1$ MeV, while the BOREXINO is surpassed for $m_{Z'} \leq 0.2$ MeV. In addition, among the constraints derived from other DM DD experiments, the WS2024 dataset provides the strongest limits, fully superseding previous bounds from PANDAX-4T and XENONnT. Although the present results do not yet reach the sensitivity of oscillation-based constraints, they remain complementary to BBN and stellar-cooling bounds in the low-mass regime.

For the $L_\mu-L_\tau$ model, the upper limits are $g_{L_\mu-L_\tau} = 3.24\times 10^{-6}$ (WS2022) and $2.42\times 10^{-6}$ (WS2024), corresponding to an improvement of approximately a factor of 1.34. These limits are stronger than the corresponding bounds derived from COHERENT and BOREXINO. As in the other $U(1)'$ scenarios, the WS2024 dataset surpasses the constraints from PandaX-4T and XENONnT throughout the explored parameter space.
The obtained limits largely exclude the low-mass portion of the parameter space favored by the $(g-2)\mu$ anomaly, while the high-mass region remains allowed. For completeness, we also show the region capable of alleviating the Hubble tension $H_0$ via $\Delta N{\text{eff}}\sim 0.2{-}0.5$ \cite{Escudero:2019gzq}, together with the bounds from SN1987A and stellar cooling, which provide complementary constraints in the sub-MeV regime.

Finally, for the $g_{L_e + 2L_\mu + 2L_\tau}$, we obtain upper limits of $1.85\times 10^{-7}$ from WS2022 and $1.47\times 10^{-7}$ from WS2024. 
Compared to other leading DD searches, the WS2024 result provides stronger constraints than both PandaX-4T and XENONnT across the entire mass range, while WS2022 remains competitive but less stringent. 
The parameter space favored by the $(g-2)_\mu$ anomaly is almost entirely excluded, leaving only a small surviving region at higher mediator masses ($m_{Z'} \gtrsim 200$ MeV). Although the present sensitivity does not yet reach the oscillation bound, the obtained limits offer valuable complementarity to those from stellar cooling and BBN considerations, particularly in the sub-MeV regime.

%%%%%%%%%%%%%%%%%%%%%
%%%%%%%%%%%%%%%%%%%%%
\begin{table*}[ht]
	\caption{The 90\% C.L. (2 d.o.f.) upper limits on the coupling constants for each light mediator model under consideration. The results are shown for the low-mass region at the scale of a few keV. For comparison, we also include the existing constraints derived from PANDAX-4T \cite{Demirci:2025qdp}, XENONnT \cite{Demirci:2025qdp, A:2022acy}, and BOREXINO \cite{Coloma:2022umy, Gninenko:2020xys}. 
	}
	\begin{center}
		\begin{ruledtabular}
		\begin{tabular}{ c c c c c c }
%			\hline
%			\hline
			\multirow{2}{1.5cm}{\textbf{Coupling}} & \multicolumn{2}{c}{\textbf{LZ (This work)}} &\multirow{2}{2.3cm}{\textbf{PANDAX-4T}} &  \multirow{2}{2cm}{\textbf{XENONnT}}
			& \multirow{2}{2cm}{\textbf{BOREXINO}}
			\\
			\cline{2-3}
			& \textbf{WS2022} &\textbf{WS2024} & & &
			\\
			\hline
%			&&&&& \\
			& & & \textit{Universal mediator models}  & &
			\\
			\hline
			$g_\phi$ & $\lesssim 6.2\times 10^{-7}$ & $ \lesssim 5.0\times 10^{-7}$ & $ \lesssim 8.3\times 10^{-7}$  & 
			$\lesssim 5.3\times 10^{-7}$  & 
			$\lesssim 1.5\times10^{-6}$ 
			\\ 
			 $g_{Z'}$ & $\lesssim 2.2\times 10^{-7}$ &  $ \lesssim 1.5\times 10^{-7}$ & $ \lesssim 2.4\times 10^{-7}$ %\cite{Demirci:2025qdp} 
             &
			$ \lesssim 1.9\times 10^{-7}$ %\cite{A:2022acy} 
            &
			$ \lesssim 6.8\times 10^{-7}$ %\cite{Coloma:2022umy}
			\\ 
			$g_{T}$ & $ \lesssim 1.2\times 10^{-7}$ & $ \lesssim 9.2\times 10^{-8}$ & $ \lesssim 1.4\times 10^{-7}$ & $\lesssim 1.0\times 10^{-7}$ &  $\cdots$ \\ \hline
		    & & & \textit{Lepton flavor-dependent $U(1)'$ models} & &  
		    \\
	   \hline
		    $g_{L_e-L_\mu}$ & $\lesssim 2.3\times 10^{-7}$ &  $ \lesssim 1.7\times 10^{-7}$ & $ \lesssim 3.0\times 10^{-7}$  &
		    $ \lesssim 2.1\times 10^{-7}$ &$ \lesssim 6.2\times 10^{-7}$ \\
		    $g_{L_e-L_\tau} $ & $\lesssim 2.3\times 10^{-7}$ &  $ \lesssim 1.6\times 10^{-7}$ & $ \lesssim 3.0\times 10^{-7}$  & $\lesssim 2.1\times 10^{-7}$  & $ \lesssim 6.4\times 10^{-7}$ \\ 
		    $g_{L_\mu-L_\tau}$ & $\lesssim 3.2\times 10^{-6}$ & $\lesssim 2.4\times 10^{-6}$ & $ \lesssim 3.3\times 10^{-6}$ & $ \lesssim 3.2\times 10^{-6}$ & $\lesssim 7.0\times 10^{-5}$  \\
		    $g_{L_e+2L_\mu+2L_\tau}$ & $\lesssim 1.9\times 10^{-7}$ & $\lesssim 1.5\times 10^{-7}$ & $ \lesssim 2.4\times 10^{-7}$& $\lesssim 1.7\times 10^{-7}$&  ... \\
            %\textbf{\multirow{-3.4}{*}{~\rotatebox[origin=c]{-90}{\parbox[r]{1.39cm}{\scriptsize This work}}}}
		\end{tabular}
			\end{ruledtabular}
	\end{center}
	\label{tab:couplings_univ}
\end{table*}
%%%%%%%%%%%%%%%%%%%%%
%%%%%%%%%%%%%%%%%%%%%
We summarize all derived constraints in Table~\ref{tab:couplings_univ}. The limits for the universal mediator scenarios can be directly read from Figs.~\ref{fig:analysis_s}, \ref{fig:analysis_v}, and \ref{fig:analysis_t}, while those for the lepton flavor-dependent of $L_e-L_\mu$, $L_e-L_\tau$, $L_\mu-L_\tau$ and $L_e+2L_\mu+2L_\tau$ models are presented in Figs.~\ref{fig:analysis_lelm}, \ref{fig:analysis_lelt}, \ref{fig:analysis_lmlt}, and \ref{fig:analysis_le2lmt}, respectively. For comparison, we also include previously published limits from PANDAX-4T, XENONnT, and BOREXINO. 
Our LZ-based constraints (WS2022 and WS2024) provide a clear improvement over the existing bounds from PANDAX-4T and XENONnT across all mediator scenarios. For the universal mediator models, the WS2024 limits improve upon the corresponding PANDAX-4T bounds by factors of 1.5–1.7 and remain about 1.1 times stronger than those from XENONnT. In the lepton-flavor–dependent $U(1)’$ models, the improvement reaches factors of 1.5–2 relative to PANDAX-4T and 1.2–1.3 relative to XENONnT. These results establish the LZ WS2024 dataset as the most sensitive probe of keV-scale mediator interactions among current direct-detection experiments.

Overall, the limits obtained in this work exhibit a substantial improvement over earlier constraints in the low-mass regime, while the recent LZ WS2024 result remains the most sensitive to date. More broadly, our analysis highlights that the LZ detector, as a state-of-the-art DM direct detection experiment, is capable of delivering highly competitive bounds that exceed many existing results due to its improved exposure and advanced detector technologies.

\section{Summary and Conclusions}\label{sec:summ}
In this study, we investigated the contributions of universal light mediator models and lepton flavor-dependent $U(1)'$ scenarios using the recent ER data released by the LZ Experiment. 
By demonstrating that these light mediators induce observable spectral distortions in solar neutrino-electron scattering, we highlight the capability of dark matter direct detection experiments to serve as sensitive probes for new physics in the sub-MeV mass regime.

By analyzing the LZ WS2022 and WS2024 datasets, we derived novel constraints on the coupling-mass parameter space. The inclusion of the WS2024 dataset significantly enhances sensitivity, yielding bounds that are substantially more stringent than those from WS2022 alone. Our results improve upon existing limits from a wide range of facilities, including stopped-pion neutrino sources, nuclear reactor, and the solar neutrino experiment, particularly in the low-mass region.
Notably, our analysis excludes a substantial portion of the parameter space favored by the muon anomalous magnetic moment, $(g-2)_\mu$, while offering complementary constraints to those from collider searches and cosmological studies.

In conclusion, we expect that the findings presented in this work will be valuable in guiding and complementing ongoing efforts to search for new physics signatures involving light mediators across various theoretical frameworks. With the continued advancement of detector technologies and increasing exposure in future DD experiments, further improvements in sensitivity are anticipated, enabling even more comprehensive exploration of light mediator parameter space.

\section*{Acknowledgments} 
This work was supported by the Scientific and Technological Research Council of Türkiye (TUBITAK) under the project no: 124F416.

\section*{Data Availability}
 The data that support the findings of this article are
 openly available \cite{our_results}.


\begin{thebibliography}{99}
%
%\cite{Davis:1968cp}
\bibitem{Davis:1968cp}
R.~Davis, Jr., D.~S.~Harmer and K.~C.~Hoffman,
%``Search for neutrinos from the sun,''
\href{http://10.1103/PhysRevLett.20.1205}{{Phys. Rev. Lett.} \textbf{20}, 1205-1209 (1968)}.

%\cite{GALLEX:1992gcp}
\bibitem{GALLEX:1992gcp}
P.~Anselmann \textit{et al.} (GALLEX Collaboration),
%``Solar neutrinos observed by GALLEX at Gran Sasso.,''
\href{https://doi.org/10.1016/0370-2693(92)91521-A}{{Phys. Lett. B} \textbf{285}, 376-389 (1992)}.
%680 citations counted in INSPIRE as of 20 Nov 2022

%\cite{Cleveland:1998nv}
\bibitem{Cleveland:1998nv}
B.~T.~Cleveland, T.~Daily, R.~Davis Jr., J.~R.~Distel, K.~Lande, C.~K.~Lee, P.~S.~Wildenhain and J.~Ullman,
%``Measurement of the solar electron neutrino flux with the Homestake chlorine detector,''
\href{http://doi.org/10.1086/305343}{{Astrophys. J.} \textbf{496}, 505-526 (1998)}.
%2867 citations counted in INSPIRE as of 20 Nov 2022

%\cite{SAGE:1999uje}
\bibitem{SAGE:1999uje}
J.~N.~Abdurashitov \textit{et al.} (SAGE Collaboration),
%``Measurement of the solar neutrino capture rate by SAGE and implications for neutrino oscillations in vacuum,''
\href{https://doi.org/10.1103/PhysRevLett.83.4686}{{Phys. Rev. Lett.} \textbf{83}, 4686-4689 (1999)}. % [arXiv:astro-ph/9907131 [astro-ph]].
%251 citations counted in INSPIRE as of 20 Nov 2022

%\cite{SNO:2002tuh}
\bibitem{SNO:2002tuh}
Q.~R.~Ahmad \textit{et al.} (SNO Collaboration),
%``Direct evidence for neutrino flavor transformation from neutral current interactions in the Sudbury Neutrino Observatory,''
\href{https://doi.org/10.1103/PhysRevLett.89.011301}{{Phys. Rev. Lett.} \textbf{89}, 011301 (2002)}.
%[arXiv:nucl-ex/0204008 [nucl-ex]].
%4217 citations counted in INSPIRE as of 20 Nov 2022

%\cite{SNO:2003bmh}
\bibitem{SNO:2003bmh}
S.~N.~Ahmed \textit{et al.} (SNO Collaboration),
%``Measurement of the total active B-8 solar neutrino flux at the Sudbury Neutrino Observatory with enhanced neutral current sensitivity,''
\href{https://doi.org/10.1103/PhysRevLett.92.181301}{{Phys. Rev. Lett.} \textbf{92}, 181301  (2004)}.
%[arXiv:nucl-ex/0309004 [nucl-ex]].
%1245 citations counted in INSPIRE as of 20 Nov 2022

%\cite{SNO:2008gqy}
\bibitem{SNO:2008gqy}
B.~Aharmim \textit{et al.} (SNO Collaboration),
%``An Independent Measurement of the Total Active B-8 Solar Neutrino Flux Using an Array of He-3 Proportional Counters at the Sudbury Neutrino Observatory,''
\href{https://doi.org/10.1103/PhysRevLett.101.111301}{{Phys. Rev. Lett.} \textbf{101}, 111301  (2008)}.
%[arXiv:0806.0989 [nucl-ex]].
%433 citations counted in INSPIRE as of 20 Nov 2022

%\cite{Kamiokande-II:1989hkh}
\bibitem{Kamiokande-II:1989hkh}
K.~S.~Hirata \textit{et al.} (Kamiokande-II Collaboration),
%``Observation of B-8 Solar Neutrinos in the Kamiokande-II Detector,''
\href{https://doi.org/10.1103/PhysRevLett.63.16}{{Phys. Rev. Lett.} \textbf{63}, 16 (1989)}.
%638 citations counted in INSPIRE as of 20 Nov 2022

%\cite{Super-Kamiokande:2001ljr}
\bibitem{Super-Kamiokande:2001ljr}
S.~Fukuda \textit{et al.} (Super-Kamiokande Collaboration),
%``Solar B-8 and hep neutrino measurements from 1258 days of Super-Kamiokande data,''
\href{https://doi.org/10.1103/PhysRevLett.86.5651}{{Phys. Rev. Lett.} \textbf{86}, 5651-5655 (2001)}.
%[arXiv:hep-ex/0103032 [hep-ex]].
%1443 citations counted in INSPIRE as of 20 Nov 2022

%\cite{Borexino:2007kvk}
\bibitem{Borexino:2007kvk}
C.~Arpesella \textit{et al.} (Borexino Collaboration),
%``First real time detection of Be-7 solar neutrinos by Borexino,''
\href{https://doi.org/10.1016/j.physletb.2007.09.054}{{Phys. Lett. B} \textbf{658}, 101-108 (2008)}.
%[arXiv:0708.2251 [astro-ph]].
%326 citations counted in INSPIRE as of 20 Nov 2022

%\cite{Borexino:2008fkj}
\bibitem{Borexino:2008fkj}
G.~Bellini \textit{et al.} (Borexino Collaboration),
%``Measurement of the solar 8B neutrino rate with a liquid scintillator target and 3 MeV energy threshold in the Borexino detector,''
\href{https://doi.org/10.1103/PhysRevD.82.033006}{{Phys. Rev. D} \textbf{82}, 033006 (2010)}.
%[arXiv:0808.2868 [astro-ph]].
%415 citations counted in INSPIRE as of 20 Nov 2022

\bibitem{tHooft:1971ucy}
G.~'t Hooft,
%``Predictions for neutrino - electron cross-sections in Weinberg's model of weak interactions,''
\href{https://doi.org/10.1016/0370-2693(71)90050-5}{Phys. Lett. B \textbf{37}, 195-196 (1971)}.

\bibitem{Bahcall:1995mm}
J.~N.~Bahcall, M.~Kamionkowski and A.~Sirlin,
%``Solar neutrinos: Radiative corrections in neutrino - electron scattering experiments,''
\href{https://doi.org/10.1103/PhysRevD.51.6146}{Phys. Rev. D \textbf{51}, 6146-6158 (1995)}.
%[arXiv:astro-ph/9502003 [astro-ph]].

%\cite{Drukier:1984vhf}
\bibitem{Drukier:1984vhf}
A.~Drukier and L.~Stodolsky,
%``Principles and Applications of a Neutral Current Detector for Neutrino Physics and Astronomy,''
\href{https://doi.org/10.1103/PhysRevD.30.2295}{{Phys. Rev. D } \textbf{30}, 2295 (1984)}.
%487 citations counted in INSPIRE as of 20 Nov 2022

%\cite{Goodman:1984dc}
\bibitem{Goodman:1984dc}
M.~W.~Goodman and E.~Witten,
%``Detectability of Certain Dark Matter Candidates,''
\href{https://doi.org/10.1103/PhysRevD.31.3059}{{Phys. Rev. D } \textbf{31}, 3059 (1985)}.
%1426 citations counted in INSPIRE as of 20 Nov 2022


%\cite{Drukier:1986tm}
\bibitem{Drukier:1986tm}
A.~K.~Drukier, K.~Freese and D.~N.~Spergel,
%``Detecting Cold Dark Matter Candidates,''
\href{https://doi.org/10.1103/PhysRevD.33.3495}{{Phys. Rev. D } \textbf{33}, 3495-3508 (1986)}.
%977 citations counted in INSPIRE as of 20 Nov 2022

%\cite{Cerdeno:2016sfi}
\bibitem{Cerdeno:2016sfi}
D.~G.~Cerde\~no, M.~Fairbairn, T.~Jubb, P.~A.~N.~Machado, A.~C.~Vincent and C.~B\oe{}hm,
%``Physics from solar neutrinos in dark matter direct detection experiments,'' 
\href{https://doi.org/10.1007/JHEP05(2016)118}{ J. High Energ. Phys. \textbf{2016}, 118 (2016)} [erratum: \href{https://doi.org/10.1007/JHEP09(2016)048}{ J. High Energ. Phys. \textbf{2016}, 48 (2016)}]. %[arXiv:1604.01025 [hep-ph]].
%108 citations counted in INSPIRE as of 20 Nov 2022

%\cite{Schwemberger:2022fjl}
\bibitem{Schwemberger:2022fjl}
T.~Schwemberger and T.~T.~Yu,
%``Detecting beyond the standard model interactions of solar neutrinos in low-threshold dark matter detectors,''
\href{https://doi.org/10.1103/PhysRevD.106.015002}{Phys. Rev. D \textbf{106}, 015002 (2022)}.
%[arXiv:2202.01254 [hep-ph]].



%\cite{XENON:2020kmp}
\bibitem{XENON:2020kmp}
E.~Aprile \textit{et al.} (XENON Collaboration),
%``Projected WIMP sensitivity of the XENONnT dark matter experiment,''
\href{http://doi.org/10.1088/1475-7516/2020/11/031}{{JCAP} \textbf{11}, 031 (2020)}. %[arXiv:2007.08796 [physics.ins-det]].
%233 citations counted in INSPIRE as of 20 Nov 2022

\bibitem{XENON:2022ltv}
E.~Aprile \textit{et al.} (XENON Collaboration),
%``Search for New Physics in Electronic Recoil Data from XENONnT,''
\href{http://doi.org/10.1103/PhysRevLett.129.161805}{Phys. Rev. Lett. \textbf{129}, 161805 (2022)}.
%[arXiv:2207.11330 [hep-ex]].

\bibitem{PandaX:2014mem}
X.~Cao \textit{et al.} (PandaX Collaboration),
%``PandaX: A Liquid Xenon Dark Matter Experiment at CJPL,''
\href{http://doi.org/10.1007/s11433-014-5521-2}{Sci. China Phys. Mech. Astron. \textbf{57}, 1476-1494 (2014)}.
%[arXiv:1405.2882 [physics.ins-det]].

%\cite{PandaX-II:2017hlx}
\bibitem{PandaX-II:2017hlx}
X.~Cui \textit{et al.} (PandaX-II Collaboration),
%``Dark Matter Results From 54-Ton-Day Exposure of PandaX-II Experiment,''
\href{http://doi.org/10.1103/PhysRevLett.119.181302}{{Phys. Rev. Lett.} \textbf{119}, 181302 (2017)}. % [arXiv:1708.06917 [astro-ph.CO]].
%981 citations counted in INSPIRE as of 20 Nov 2022

\bibitem{PandaX:2024cic} X.~Zeng \textit{et al.} (PandaX Collaboration),
    %``Exploring New Physics with PandaX-4T Low Energy Electronic Recoil Data,''
       \href{https://doi.org/10.1103/PhysRevLett.134.041001}{Phys. Rev. Lett. \textbf{134}, 041001 (2025)}.

%\cite{LUX:2015abn}
\bibitem{LUX:2015abn}
D.~S.~Akerib \textit{et al.} (LUX-ZEPLIN Collaboration),
%``Improved Limits on Scattering of Weakly Interacting Massive Particles from Reanalysis of 2013 LUX Data,''
\href{http://doi.org/10.1103/PhysRevLett.116.161301}{{Phys. Rev. Lett.} \textbf{116}, 161301  (2016)}.% [arXiv:1512.03506 [astro-ph.CO]].
%549 citations counted in INSPIRE as of 20 Nov 2022

%\cite{LZ:2018qzl}
\bibitem{LZ:2018qzl}
D.~S.~Akerib \textit{et al.} (LUX-ZEPLIN Collaboration),
%``Projected WIMP sensitivity of the LUX-ZEPLIN dark matter experiment,''
\href{http://doi.org/10.1103/PhysRevD.101.052002}{{Phys. Rev. D} \textbf{101}, 052002 (2020)}. % [arXiv:1802.06039 [astro-ph.IM]].
%348 citations counted in INSPIRE as of 20 Nov 2022

\bibitem{LZ:2022lsv}
J.~Aalbers \textit{et al.} (LUX-ZEPLIN Collaboration),
%``First Dark Matter Search Results from the LUX-ZEPLIN (LZ) Experiment,''
\href{https://doi.org/10.1103/PhysRevLett.131.041002}{{Phys. Rev. Lett. } \textbf{131}, 041002 (2023)}.
%[arXiv:2207.03764 [hep-ex]].

\bibitem{LZ:2024zvo}
J.~Aalbers \textit{et al.} (LUX-ZEPLIN Collaboration),
%``Dark Matter Search Results from 4.2{\,}{\,}Tonne-Years of Exposure of the LUX-ZEPLIN (LZ) Experiment,''
\href{https://doi.org/10.1103/4dyc-z8zf}{{Phys. Rev. Lett. } \textbf{135}, 011802 (2025)}.
%[arXiv:2410.17036 [hep-ex]].

\bibitem{LZ:2025zpw}
D.~S.~Akerib \textit{et al.} (LUX-ZEPLIN Collaboration),
%``Search for New Physics via Low-Energy Electron Recoils with a 4.2 Tonne{\texttimes} Year Exposure from the LZ Experiment,''
[arXiv:2511.17350 [hep-ex]].

\bibitem{DARKSIDE20K:2021}
V.~Pesudo \textit{et al.} (DarkSide-20k Collaboration),
%``Measurement of the underground argon radiopurity for Dark Matter direct searches,''
\href{http://doi.org/10.1088/1742-6596/2156/1/012043}{ J. Phys.: Conf. Ser. \textbf{2156}, 012043  (2021)}.

\bibitem{DARWIN:2020bnc}
J.~Aalbers \textit{et al.} (DARWIN Collaboration),
%``Solar neutrino detection sensitivity in DARWIN via electron scattering,''
\href{http://doi.org/10.1140/epjc/s10052-020-08602-7}{Eur. Phys. J. C \textbf{80}, 1133  (2020)}.
%[arXiv:2006.03114 [physics.ins-det]].

%\cite{Aalbers:2022dzr}
\bibitem{Aalbers:2022dzr}
J.~Aalbers \textit{et al.},
%``A next-generation liquid xenon observatory for dark matter and neutrino physics,''
\href{http://doi.org/10.1088/1361-6471/ac841a}{J. Phys. G \textbf{50}, 013001 (2023)}.
%194 citations counted in INSPIRE as of 07 Dec 2025

\bibitem{Cirelli:2013ufw}
M.~Cirelli, E.~Del Nobile and P.~Panci,
%``Tools for model-independent bounds in direct dark matter searches,''
\href{http://doi.org/10.1088/1475-7516/2013/10/019}{JCAP \textbf{10}, 019  (2013)}.
%[arXiv:1307.5955 [hep-ph]].

%\cite{Abdallah:2015ter}
\bibitem{Abdallah:2015ter}
J.~Abdallah, H.~Araujo, A.~Arbey, A.~Ashkenazi, A.~Belyaev, J.~Berger, C.~Boehm, A.~Boveia, A.~Brennan and J.~Brooke,
%``Simplified Models for Dark Matter Searches at the LHC,''
\href{https://doi.org/10.1016/j.dark.2015.08.001}{{Phys. Dark Univ.} \textbf{9-10}, 8-23 (2015)}. %[arXiv:1506.03116 [hep-ph]].
%381 citations counted in INSPIRE as of 20 Nov 2022

\bibitem{Mohapatra:1980qe}
R.~N.~Mohapatra and R.~E.~Marshak,
%``Local B-L Symmetry of Electroweak Interactions, Majorana Neutrinos and Neutron Oscillations,''
\href{https://doi.org/10.1103/PhysRevLett.44.1316}{Phys. Rev. Lett. \textbf{44}, 1316-1319  (1980)	[erratum: Phys. Rev. Lett. \textbf{44}, 1643 (1980)]}. 

%\cite{He:1991qd}
\bibitem{He:1991qd}
X.~G.~He, G.~C.~Joshi, H.~Lew and R.~R.~Volkas,
%``Simplest Z-prime model,''
\href{https://doi.org/10.1103/PhysRevD.44.2118}{Phys. Rev. D \textbf{44}, 2118-2132 (1991)}.
%426 citations counted in INSPIRE as of 28 Sep 2024

\bibitem{ParticleDataGroup:2024cfk}
S.~Navas \textit{et al.} (Particle Data Group),
%``Review of particle physics,''
\href{https://doi.org/10.1103/PhysRevD.110.030001}{Phys. Rev. D \textbf{110}, 030001 (2024)}.

\bibitem{Erler:2013c}
J. Erler and S. Su,
%``The weak neutral current,''
\href{https://doi.org/10.1016/j.ppnp.2013.03.004}{Prog. Part. Nucl. Phys. \textbf{71}, 119 (2013)}.

%\cite{AtzoriCorona:2025ygn}
\bibitem{AtzoriCorona:2025ygn}
M.~Atzori Corona, M.~Cadeddu, N.~Cargioli, F.~Dordei and C.~Giunti,
%``Reactor antineutrinos CE{\ensuremath{\nu}}NS on germanium: CONUS+ and TEXONO as a new gateway to SM and BSM physics,''
\href{http://dx.doi.org/10.1103/n563-8v8d}{Phys. Rev. D \textbf{112}, 015007 (2025)}.
%8 citations counted in INSPIRE as of 30 Jul 2025

\bibitem{Formaggio:2012cpf}
J.~A.~Formaggio and G.~P.~Zeller,
%``From eV to EeV: Neutrino Cross Sections Across Energy Scales,''
\href{https://doi.org/10.1103/RevModPhys.84.1307}{Rev. Mod. Phys. \textbf{84}, 1307-1341 (2012)}.
%[arXiv:1305.7513 [hep-ex]].

%\cite{Allanach:2019}
\bibitem{Allanach:2019} B. C. Allanach, J. Davighi and S. Melville, 
%An anomaly-free ATLAS: charting the space of flavour-dependent gauged U(1) extensions of the Standard Model, 
\href{https://doi.org/10.1007/JHEP02(2019)082}{ J. High Energ. Phys. \textbf{2019}, 82 (2019)} [erratum: \href{https://doi.org/10.1007/JHEP08(2019)064}{ J. High Energ. Phys. \textbf{2019}, 64 (2019)}].

\bibitem{Ballet:2019}
P. Ballett, M. Hostert, S. Pascoli,  Y. F. Perez-Gonzalez, Z. Tabrizi, and  R. Z. Funchal,
 %${Z}^{\ensuremath{'}}\mathrm{s}$ in neutrino scattering at DUNE,
\href{https://doi.org/10.1103/PhysRevD.100.055012}{Phys. Rev. D \textbf{100}, 055012 (2019)}.

\bibitem{DeRomeri:2022twg}
V.~De Romeri, O.~G.~Miranda, D.~K.~Papoulias, G.~Sanchez Garcia, M.~T\'ortola and J.~W.~F.~Valle,
%``Physics implications of a combined analysis of COHERENT CsI and LAr data,''
\href{https://doi.org/10.1007/JHEP04(2023)035}{J. High Energ. Phys. \textbf{2023}, 35 (2023)}. 
%[arXiv:2211.11905 [hep-ph]].

\bibitem{Demirci:2023tui}
M.~Demirci and M.~F.~Mustamin,
%``Solar neutrino constraints on light mediators through coherent elastic neutrino-nucleus scattering,''
\href{https://doi.org/10.1103/PhysRevD.109.015021}{Phys. Rev. D \textbf{109}, 015021 (2024)}.
%[arXiv:2312.17502 [hep-ph]].

\bibitem{Barranco:2011wx}
J.~Barranco, A.~Bolanos, E.~A.~Garces, O.~G.~Miranda and T.~I.~Rashba,
%``Tensorial NSI and Unparticle physics in neutrino scattering,''
\href{https://doi.org/10.1142/S0217751X12501473}{{Int. J. Mod. Phys. A} \textbf{27}, 1250147 (2012)}. %[arXiv:1108.1220 [hep-ph]].

%\cite{AtzoriCorona:2022moj}
\bibitem{AtzoriCorona:2022moj}
M.~Atzori Corona, M.~Cadeddu, N.~Cargioli, F.~Dordei, C.~Giunti, Y.~F.~Li, E.~Picciau, C.~A.~Ternes and Y.~Y.~Zhang,
%``Probing light mediators and (g \ensuremath{-} 2)$_{μ}$ through detection of coherent elastic neutrino nucleus scattering at COHERENT,''
\href{http://doi.org/10.1007/JHEP05(2022)109}{J. High Energ. Phys. \textbf{2022}, 109 (2022)}.
%[arXiv:2202.11002 [hep-ph]].
%31 citations counted in INSPIRE as of 06 Sep 2023

%\cite{Coloma:2022umy}
\bibitem{Coloma:2022umy}
P.~Coloma, M.~C.~Gonzalez-Garcia, M.~Maltoni, J.~P.~Pinheiro and S.~Urrea,
%``Constraining new physics with Borexino Phase-II spectral data,''
\href{http://doi.org/10.1007/JHEP07(2022)138}{J. High Energ. Phys. \textbf{2022}, 138 (2022)}. %[arXiv:2204.03011 [hep-ph]].
%6 citations counted in INSPIRE as of 07 Nov 2022

\bibitem{Gninenko:2020xys}
S.~Gninenko and D.~Gorbunov,
%``Refining constraints from Borexino measurements on a light Z'-boson coupled to L\ensuremath{\mu}-L\ensuremath{\tau} current,''
\href{http://doi.org/10.1016/j.physletb.2021.136739}{Phys. Lett. B \textbf{823}, 136739 (2021)}.
%[arXiv:2007.16098 [hep-ph]].

%\cite{Demirci:2025qdp}
\bibitem{Demirci:2025qdp}
M.~Demirci and M.~F.~Mustamin,
%``Probing light mediators with recent PandaX-4T low-energy electron recoil data,''
\href{https://doi.org/10.1103/PhysRevD.111.055032}{Phys. Rev. D \textbf{111}, no.5, 055032 (2025)}.
%[arXiv:2502.20026 [hep-ph]].
%4 citations counted in INSPIRE as of 02 Dec 2025

\bibitem{DeRomeri:2024dbv}
V.~De Romeri, D.~K.~Papoulias and C.~A.~Ternes,
%``Light vector mediators at direct detection experiments,''
\href{https://doi.org/10.1007/JHEP05(2024)165}{ J. High Energ. Phys. \textbf{2024}, 165 (2024)}.
%[arXiv:2402.05506 [hep-ph]].

\bibitem{Bauer:2018onh}
M.~Bauer, P.~Foldenauer and J.~Jaeckel,
%``Hunting All the Hidden Photons,''
\href{https://doi.org/10.1007/JHEP07(2018)094}{J. High Energ. Phys. \textbf{2018}, 94 (2018)}.
%[arXiv:1803.05466 [hep-ph]].

\bibitem{Altmannshofer:2019}
W. Altmannshofer, S. Gori, J. Mart\'{\i}n-Albo, A. Sousa, and M. Wallbank,
%``Neutrino tridents at DUNE,''
\href{https://doi.org/10.1103/PhysRevD.100.115029}{Phys. Rev. D \textbf{100}, 115029 (2019)}.

\bibitem{Chen:2016eab}
J.~W.~Chen, H.~C.~Chi, C.~P.~Liu and C.~P.~Wu,
%``Low-energy electronic recoil in xenon detectors by solar neutrinos,''
\href{https://doi.org/10.1016/j.physletb.2017.10.029}{Phys. Lett. B \textbf{774}, 656-661 (2017)}.
%[arXiv:1610.04177 [hep-ex]].

\bibitem{xraydata:2009}
A.~Thompson \textit{et al.},
%``Review of particle physics,''
\href{https://xdb.lbl.gov/}{X-ray data booklet (2009)}.

\bibitem{Kouzakov:2017hbc}
K.~A.~Kouzakov and A.~I.~Studenikin,
%``Electromagnetic properties of massive neutrinos in low-energy elastic neutrino-electron scattering,''
\href{https://doi.org/10.1103/PhysRevD.95.055013}{Phys. Rev. D \textbf{95}, 055013 (2017) [erratum: Phys. Rev. D \textbf{96}, 099904 (2017)]}.
%[arXiv:1703.00401 [hep-ph]].

\bibitem{Hsieh:2019hug}
C.~C.~Hsieh, L.~Singh, C.~P.~Wu, J.~W.~Chen, H.~C.~Chi, C.~P.~Liu, M.~K.~Pandey and H.~T.~Wong,
%``Discovery potential of multiton xenon detectors in neutrino electromagnetic properties,''
\href{https://doi.org/10.1103/PhysRevD.100.073001}{Phys. Rev. D \textbf{100}, 073001 (2019)}.
%[arXiv:1903.06085 [hep-ph]].

\bibitem{Maltoni:2015kca}
M.~Maltoni and A.~Y.~Smirnov,
%``Solar neutrinos and neutrino physics,''
\href{https://doi.org/10.1140/epja/i2016-16087-0}{Eur. Phys. J. A \textbf{52} no.4, 87 (2016)}.
%[arXiv:1507.05287 [hep-ph]].

\bibitem{Esteban:2024eli}
I.~Esteban, M.~C.~Gonzalez-Garcia, M.~Maltoni, I.~Martinez-Soler, J.~P.~Pinheiro and T.~Schwetz,
%``NuFit-6.0: updated global analysis of three-flavor neutrino oscillations,''
\href{http://dx.doi.org/10.1007/JHEP12(2024)216}{{J. High Energ. Phys}  \textbf{12}, 216 (2024)}.
%[arXiv:2410.05380 [hep-ph]].

\bibitem{Pereira:2023rte}
G.~Pereira \textit{et al.} [LZ],
%``Energy resolution of the LZ detector for high-energy electronic recoils,''
\href{http://dx.doi.org/10.1088/1748-0221/18/04/C04007}{JINST \textbf{18} (2023), C04007}.


\bibitem{Baker:1983tu}
S.~Baker and R.~D.~Cousins,
%``Clarification of the Use of Chi Square and Likelihood Functions in Fits to Histograms,''
\href{http://dx.doi.org/10.1016/0167-5087(84)90016-4}{Nucl. Instrum. Meth. \textbf{221}, 437-442 (1984)}.

%\cite{Fogli:2002pt}
\bibitem{Fogli:2002pt}
G.~L.~Fogli, E.~Lisi, A.~Marrone, D.~Montanino and A.~Palazzo,
%``Getting the most from the statistical analysis of solar neutrino oscillations,''
\href{http://doi.org/10.1103/PhysRevD.66.053010}{{Phys. Rev. D} \textbf{66}, 053010 (2002)}.
%[arXiv:hep-ph/0206162 [hep-ph]].
%405 citations counted in INSPIRE as of 01 Jun 2023

\bibitem{Baxter:2021pqo}
D.~Baxter, I.~M.~Bloch, E.~Bodnia, X.~Chen, J.~Conrad, P.~Di Gangi, J.~E.~Y.~Dobson, D.~Durnford, S.~J.~Haselschwardt and A.~Kaboth, \textit{et al.}
%``Recommended conventions for reporting results from direct dark matter searches,''
\href{http://doi.org/10.1140/epjc/s10052-021-09655-y}{Eur. Phys. J. C \textbf{81}, 907 (2021)}.
%[arXiv:2105.00599 [hep-ex]].

\bibitem{Bahcall:1989ks}
J.~N.~Bahcall, Neutrino Astrophysics, \href{https://www.cambridge.org/tr/universitypress/subjects/physics/astrophysics/neutrino-astrophysics}{Cambridge University Press (1989), ISBN: 9780521379755}.
%``NEUTRINO ASTROPHYSICS,''

%\cite{Vinyoles:2016djt}
\bibitem{Vinyoles:2016djt}
N.~Vinyoles, A.~M.~Serenelli, F.~L.~Villante, S.~Basu, J.~Bergstr\"om, M.~C.~Gonzalez-Garcia, M.~Maltoni, C.~Pe\~na-Garay and N.~Song,
%``A new Generation of Standard Solar Models,''
\href{https://doi.org/10.3847/1538-4357/835/2/202}{{Astrophys. J.} \textbf{835}, 202 (2017)}.
%[arXiv:1611.09867 [astro-ph.SR]].
%230 citations counted in INSPIRE as of 20 Nov 2022

\bibitem{Lindner:2020kko}
M.~Lindner, Y.~Mambrini, T.~B.~de Melo and F.~S.~Queiroz,
%``XENON1T anomaly: A light Z' from a Two Higgs Doublet Model,''
\href{http://doi.org/10.1016/j.physletb.2020.135972}{Phys. Lett. B \textbf{811}, 135972  (2020)}.
%[arXiv:2006.14590 [hep-ph]].

\bibitem{Bilmis:2015lja}
S.~Bilmis, I.~Turan, T.~M.~Aliev, M.~Deniz, L.~Singh and H.~T.~Wong,
%``Constraints on Dark Photon from Neutrino-Electron Scattering Experiments,''
\href{http://doi.org/10.1103/PhysRevD.92.033009}{Phys. Rev. D \textbf{92}, 033009 (2015)}.
%[arXiv:1502.07763 [hep-ph]].

\bibitem{Coloma:2022avw}
P.~Coloma, I.~Esteban, M.~C.~Gonzalez-Garcia, L.~Larizgoitia, F.~Monrabal and S.~Palomares-Ruiz,
%``Bounds on new physics with data of the Dresden-II reactor experiment and COHERENT,''
\href{http://doi.org/10.1007/JHEP05(2022)037}{J. High Energ. Phys. \textbf{2022}, 37 (2022)}.
%[arXiv:2202.10829 [hep-ph]].

\bibitem{CONNIE:2019xid}
A.~Aguilar-Arevalo \textit{et al.} (CONNIE Collaboration),
%``Search for light mediators in the low-energy data of the CONNIE reactor neutrino experiment,''
\href{http://doi.org/10.1007/JHEP04(2020)054}{J. High Energ. Phys. \textbf{2020}, 54 (2020)}.
%[arXiv:1910.04951 [hep-ex]].

\bibitem{CONUS:2021dwh}
H.~Bonet \textit{et al.} (CONUS Collaboration),
%``Novel constraints on neutrino physics beyond the standard model from the CONUS experiment,''
\href{https://doi.org/10.1007/JHEP05(2022)085}{ J. High Energ. Phys. \textbf{2022}, 85 (2022)}. 
%[arXiv:2110.02174 [hep-ph]].

\bibitem{Chattaraj:2025}
A. Chattaraj, A. Majumdar and R. Srivastava, Probing Standard Model and Beyond with Reactor CE$\nu$NS Data of CONUS+ experiment, 
\href{https://doi.org/10.48550/arXiv.2501.12441}{arXiv:2501.12441 [hep-ph]}.

%\cite{Melas:2023olz} DUNE
\bibitem{Melas:2023olz}
P.~Melas, D.~K.~Papoulias and N.~Saoulidou,
%``Probing generalized neutrino interactions with the DUNE Near Detector,''
\href{http://doi.org/10.1007/JHEP07(2023)190}{ J. High Energ. Phys. \textbf{2023}, 190 (2023)}.
%[arXiv:2303.07094 [hep-ph]].
%3 citations counted in INSPIRE as of 15 Sep 2023

%\cite{A:2022acy}
\bibitem{A:2022acy}
S.~K.~A., A.~Majumdar, D.~K.~Papoulias, H.~Prajapati and R.~Srivastava,
%``Implications of first LZ and XENONnT results: A comparative study of neutrino properties and light mediators,''
\href{http://doi.org/10.1016/j.physletb.2023.137742}{{Phys. Lett. B} \textbf{839}, 137742 (2023)}.
%[arXiv:2208.06415 [hep-ph]].
%20 citations counted in INSPIRE as of 23 Sep 2023

%\cite{NA64:2022yly}
\bibitem{NA64:2022yly}
Y.~M.~Andreev \textit{et al.} (NA64 Collaboration),
%``Search for a New B-L Z' Gauge Boson with the NA64 Experiment at CERN,''
\href{http://doi.org/10.1103/PhysRevLett.129.161801}{{Phys. Rev. Lett.} \textbf{129}, 161801  (2022)}.
%[arXiv:2207.09979 [hep-ex]].
%5 citations counted in INSPIRE as of 02 Jan 2023

\bibitem{A1:2011yso}
H.~Merkel \textit{et al.} (A1 Collaboration),
%``Search for Light Gauge Bosons of the Dark Sector at the Mainz Microtron,''
\href{http://doi.org/10.1103/PhysRevLett.106.251802}{Phys. Rev. Lett. \textbf{106}, 251802 (2011)}.
%[arXiv:1101.4091 [nucl-ex]].

\bibitem{ALICE:2012aqc}
B.~Abelev \textit{et al.} (ALICE Collaboration),
%``Centrality Dependence of Charged Particle Production at Large Transverse Momentum in Pb--Pb Collisions at $\sqrt{s_{\rm{NN}}} = 2.76$ TeV,''
\href{http://doi.org/10.1016/j.physletb.2013.01.051}{Phys. Lett. B \textbf{720}, 52-62 (2013)}.
%[arXiv:1208.2711 [hep-ex]].

\bibitem{BaBar:2014zli}
J.~P.~Lees \textit{et al.} (BaBar Collaboration),
%``Search for a Dark Photon in $e^+e^-$ Collisions at BaBar,''
\href{http://doi.org/10.1103/PhysRevLett.113.201801}{Phys. Rev. Lett. \textbf{113}, 201801 (2014)}.
%[arXiv:1406.2980 [hep-ex]].

\bibitem{PHENIX:2014duq}
A.~Adare \textit{et al.} (PHENIX Collaboration),
%``Search for dark photons from neutral meson decays in $p + p$ and $d$ + Au collisions at $\sqrt{s_{NN}} =$ 200 GeV,''
\href{http://doi.org/10.1103/PhysRevC.91.031901}{Phys. Rev. C \textbf{91}, 031901 (2015)}.
%[arXiv:1409.0851 [nucl-ex]].

\bibitem{NA482:2015wmo}
J.~R.~Batley \textit{et al.} (NA48/2 Collaboration),
%``Search for the dark photon in $\pi^0$ decays,''
\href{http://doi.org/10.1016/j.physletb.2015.04.068}{Phys. Lett. B \textbf{746}, 178-185 (2015)}.
%[arXiv:1504.00607 [hep-ex]].

\bibitem{Coloma:2020gfv}
P.~Coloma, M.~C.~Gonzalez-Garcia and M.~Maltoni,
%``Neutrino oscillation constraints on U(1)' models: from non-standard interactions to long-range forces,''
\href{https://doi.org/10.1007/JHEP01(2021)114}{ J. High Energ. Phys. \textbf{2021}, 114 (2021)}	\href{https://doi.org/10.1007/JHEP11(2022)115}{[erratum: J. High Energ. Phys. \textbf{2022}, 115 (2022)]}.
%[arXiv:2009.14220 [hep-ph]].

\bibitem{Muong-2:2023cdq}
D.~P.~Aguillard \textit{et al.} (Muon g-2 Collaboration),
%``Measurement of the Positive Muon Anomalous Magnetic Moment to 0.20~ppm,''
\href{http://doi.org/10.1103/PhysRevLett.131.161802}{Phys. Rev. Lett. \textbf{131}, 161802 (2023)}.
%[arXiv:2308.06230 [hep-ex]].

%\cite{Blinov:2019gcj}
\bibitem{Blinov:2019gcj}
N.~Blinov, K.~J.~Kelly, G.~Z.~Krnjaic and S.~D.~McDermott,
%``Constraining the Self-Interacting Neutrino Interpretation of the Hubble Tension,''
\href{http://doi.org/10.1103/PhysRevLett.123.191102}{{Phys. Rev. Lett.} \textbf{123}, 191102  (2019)}.
%[arXiv:1905.02727 [astro-ph.CO]].
%163 citations counted in INSPIRE as of 22 Sep 2023

\bibitem{Huang:2018}
G.-y. Huang, T. Ohlsson, and S. Zhou,
%``Observational constraints on secret neutrino interactions from big bang nucleosynthesis,"
\href{https://link.aps.org/doi/10.1103/PhysRevD.97.075009}{{Phys. Rev. D} \textbf{97}, 075009  (2018)}.

\bibitem{Heurtier:2017}
L. Heurtier and Y. Zhang, %"Supernova constraints on massive (pseudo)scalar coupling to neutrinos",
\href{http://doi.org/10.1088/1475-7516/2017/02/042}{{JCAP} \textbf{02}, 042  (2017)}.

\bibitem{Croon:2020lrf}
D.~Croon, G.~Elor, R.~K.~Leane and S.~D.~McDermott,
%``Supernova Muons: New Constraints on $Z$' Bosons, Axions and ALPs,''
\href{http://doi.org/10.1007/JHEP01(2021)107}{ J. High Energ. Phys. \textbf{2021}, 107 (2021)}.
%[arXiv:2006.13942 [hep-ph]].

\bibitem{Escudero:2019gzq}
M.~Escudero, D.~Hooper, G.~Krnjaic and M.~Pierre,
%``Cosmology with A Very Light L$_{\mu}$ \ensuremath{-} L$_{\tau}$ Gauge Boson,''
\href{http://doi.org/10.1007/JHEP03(2019)071}{J. High Energ. Phys. \textbf{2019}, 71 (2019)}.
%[arXiv:1901.02010 [hep-ph]].

\bibitem{LiXu:2023}
S.-P. Li and X.-J. Xu, %"Production rates of dark photons and Z' in the Sun and stellar cooling bounds",
\href{https://dx.doi.org/10.1088/1475-7516/2023/09/009}{JCAP \textbf{09}, 009 (2023)}.

\bibitem{our_results}
M. Demirci and M. F. Mustamin, “Data Analysis Results for: Solar Neutrino Probes of Light New Physics: Updated Limits from LUX-ZEPLIN Experiment”. Zenodo, 2026, \href{https://doi.org/10.5281/zenodo.18462143}{doi: 10.5281/zenodo.18462143}.

\end{thebibliography}
\end{document}